\documentclass[aps,pra,twocolumn,showkeys,amssymb,floatfix,longbibliography,superscriptaddress]{revtex4-2}

\usepackage{graphicx}
\usepackage{amsmath,amsfonts,amssymb}
\usepackage[hypertexnames=false]{hyperref}
\usepackage{braket}
\usepackage[svgnames]{xcolor}
\usepackage{color}
\usepackage{enumerate}
\usepackage[caption=false]{subfig}
\usepackage{multirow}
\usepackage{qcircuit}
\usepackage{placeins}
\usepackage{url}
\usepackage{dsfont}
\usepackage[multidot]{grffile}
\usepackage{textcomp}

\begin{document}

\title{Large-Scale Simulation of Shor's Quantum Factoring Algorithm}

\author{Dennis Willsch}
\thanks{Corresponding author: Dennis Willsch}
\email{d.willsch@fz-juelich.de}
\affiliation{Jülich Supercomputing Centre, Institute for Advanced Simulation, Forschungszentrum Jülich, 52425 Jülich, Germany}
\author{Madita Willsch}
\affiliation{Jülich Supercomputing Centre, Institute for Advanced Simulation, Forschungszentrum Jülich, 52425 Jülich, Germany}
\affiliation{AIDAS, 52425 J\"ulich, Germany}
\author{Fengping Jin}
\affiliation{Jülich Supercomputing Centre, Institute for Advanced Simulation, Forschungszentrum Jülich, 52425 Jülich, Germany}
\author{Hans De Raedt}
\affiliation{Jülich Supercomputing Centre, Institute for Advanced Simulation, Forschungszentrum Jülich, 52425 Jülich, Germany}
\affiliation{Zernike Institute for Advanced Materials, University of Groningen, Nijenborgh 4, 9747 AG Groningen, The Netherlands}
\author{Kristel Michielsen}
\affiliation{Jülich Supercomputing Centre, Institute for Advanced Simulation, Forschungszentrum Jülich, 52425 Jülich, Germany}
\affiliation{AIDAS, 52425 J\"ulich, Germany}
\affiliation{Department of Physics, RWTH Aachen University, 52056 Aachen, Germany}

\date{\today}

\begin{abstract}
   \textbf{Abstract:} Shor's factoring algorithm is one of the most anticipated applications of quantum computing. 
   However, the limited capabilities of today's quantum computers only permit a study of Shor's algorithm for very small numbers.
   Here we show how large GPU-based supercomputers can be used to assess the performance of Shor's algorithm for numbers that are out of reach for current and near-term quantum hardware.
   First, we study Shor's original factoring algorithm. While theoretical bounds suggest success probabilities of only $3$--$4\,\%$,
   we find average success probabilities above $50\,\%$, due to a high frequency of ``lucky'' cases, defined as successful factorizations despite unmet sufficient conditions.
   Second, we investigate a powerful post-processing procedure, by which the success probability can be brought arbitrarily close to one, with only a single run of Shor's quantum algorithm.
   Finally, we study the effectiveness of this post-processing procedure in the presence of typical errors in quantum processing hardware.
   We find that the quantum factoring algorithm exhibits a particular form of universality and resilience against the different types of errors.
   The largest semiprime that we have factored by executing Shor's algorithm on a GPU-based supercomputer, without exploiting prior knowledge of the solution, is $549755813701=712321\times771781$.
   We put forward the challenge of factoring, without oversimplification, a non-trivial semiprime larger than this number on any quantum computing device.
\end{abstract}


\maketitle

\section{Introduction}

The challenge of factoring integers is one of the oldest problems in mathematics~\cite{Bressoud,Lehman1974FactoringLargeIntegers}.
Famous mathematicians such as Fermat, Euler, and Gauss have made substantial contributions to the problem, and even algorithms discovered by the ancient Greeks---the Euclidean algorithm and the sieve of Eratosthenes---are still in use today.
The state-of-the-art algorithms are based on the general number field sieve~\cite{lenstra1993developmentOfTheNumberFieldSieve} and have recently achieved the factorization of RSA-250 from the famous RSA factoring challenge~\cite{Boudot2020FactoringRSA250}. 
Still, all known algorithms exhibit at best subexponential time and space complexity~\cite{Kleinjung2010FactorizeRSA768Modulus,Boudot2020FactoringRSA250}.
The difficulty of solving this type of problem on classical computers is an integral aspect of modern data and communication security~\cite{Gidney2021HowToFactor2048RSAShor,Biasse2023QuantumAlgorithmsForAttackingHardnessAssumptionsCryptography}.

In 1994, Peter Shor proposed an algorithm to factor integers on quantum computers with an exponential speedup~\cite{shor1994factoring,shor1994factoring,ekert1996quantumalgorithms,shor1997algorithm} over the best known classical algorithms. Factoring an $L$-bit integer $N$ with the conventional Shor algorithm~\cite{shor1997algorithm} requires at least $3L$ qubits: $L=\lfloor\log_2N\rfloor+1$ qubits to represent $N$, and $t=\lceil2\log_2N\rceil\approx2L$ qubits for the Quantum Fourier Transform (QFT), plus $O(L)$ qubits for the modular exponentiation~\cite{VanMeter2005FastQuantumModularExponentiationShor,Gidney2021HowToFactor2048RSAShor}. Kitaev, Griffiths, and Niu realized that by replacing the QFT with a semiclassical Fourier transform, only a single qubit can be reused $t$ times to obtain the same result~\cite{Kitaev1995MeasurementAbelianStabilizerProblem, Griffiths1996SemiclassicalQuantumFourierTransform} (also known as \emph{qubit recycling}~\cite{Parker2000EfficientShorSingleQubitAndLogNMixedQubits,MartinLopez2012ShorExperimentQubitRecycling} or \emph{dynamic quantum computing}~\cite{corcoles2021exploitingDynamicQuantumCircuits}). It is thus possible to run Shor's algorithm with only $L+1$ qubits to factor $L$-bit integers (which is less than required by the best adiabatic algorithm~\cite{Peng2008QuantumAdiabaticAlgorithmForFactorization,Hegade2021DigitizedAdiabaticQuantumFactorization}). We refer to this variant as the \emph{iterative Shor algorithm}.

The iterative Shor algorithm has been executed on real quantum computing devices to factor 15, 21, and 35~\cite{MartinLopez2012ShorExperimentQubitRecycling,Monz2016ShorOnTrappedIons,Amico2019ShorOnIBMQ}, without relying on oversimplification~\cite{Smolin2013OversimplifyingShorFactoring}. 
Implementing the algorithm for integers beyond 35 continues to pose substantial experimental challenges~\cite{Gidney2021HowToFactor2048RSAShor,Gouzien2021ShorFactoring2048RSAIn177DaysWith13436Qubits}.

\begin{figure*}
  \centering
  \includegraphics[width=\textwidth]{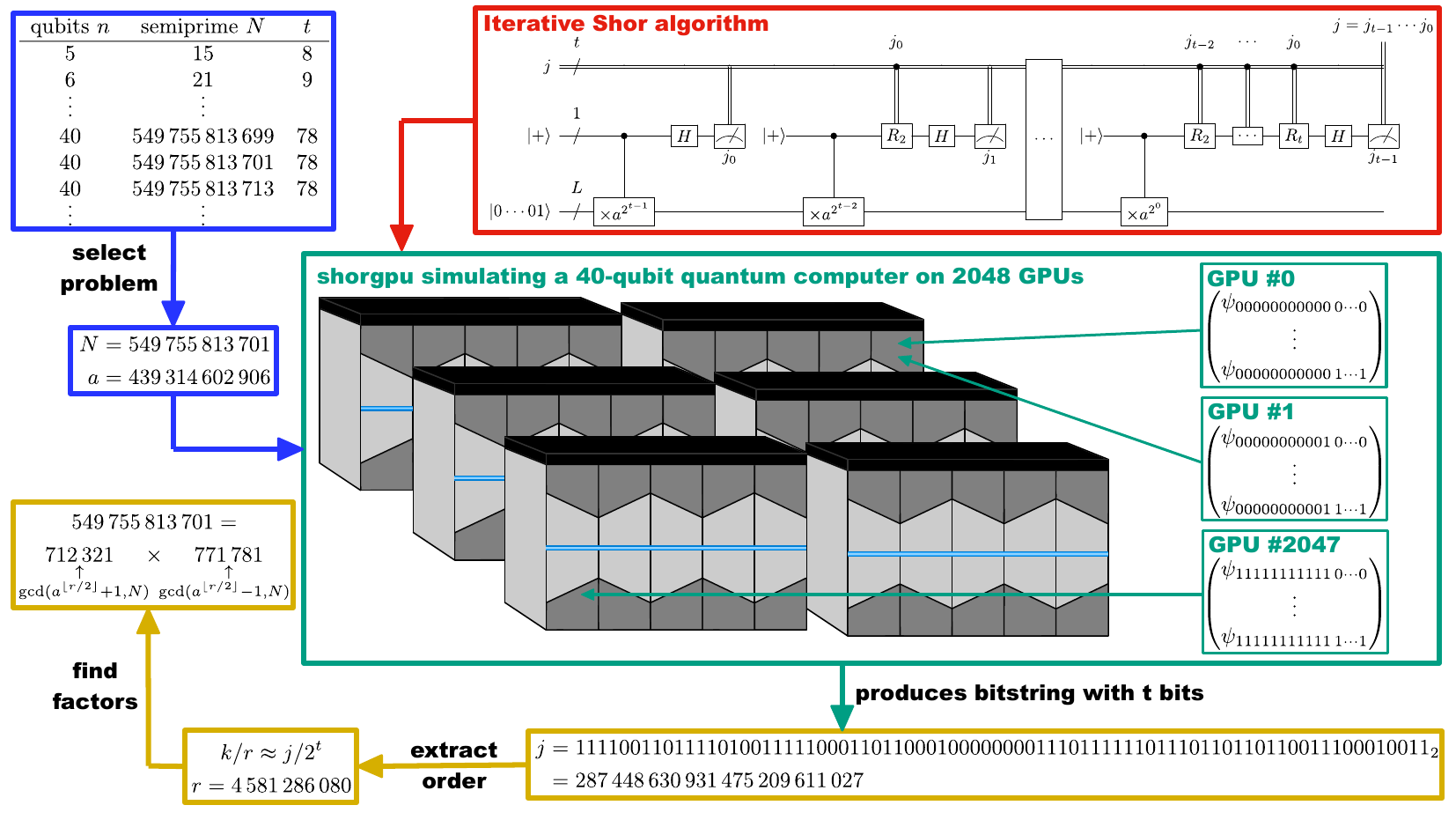}
  \caption{\textbf{Scheme to test Shor's algorithm.} 
  After selecting an $L$-bit semiprime $N=p\times q$ to factor and a random integer $1<a<N$ coprime to $N$ (blue), a quantum computer or quantum computer simulator with $n=L+1$ qubits runs the iterative Shor algorithm (red) and produces several bitstrings $j$ with $t$ bits (green). 
  Here, ``iterative'' means that one qubit is measured and reused $t$ times to produce the $t$ classical bits of each $j$.
  Every bitstring $j$ is analyzed using either Shor's~\cite{shor1994factoring,ekert1996quantumalgorithms,shor1997algorithm,NielsenChuang} or Eker\aa's~\cite{Ekera2021OnCompletelyFactoringAnyIntegerShor,Ekera2022OnTheSuccessProbabilityOfQuantumOrderFindingShor} post-processing method (yellow), independent of whether certain algorithmic requirements on $j$ are satisfied or not.
  Here, $k$ ($r$) denotes the numerator (denominator) obtained from a continued fraction expansion of $j/2^t$.
  Note that the expression for the factors in the yellow section is specific to Shor's post-processing; for Eker\aa's post-processing, $\mathrm{gcd}(a^{\lfloor r/2\rfloor}\pm1,N)$ has to be replaced by $\mathrm{gcd}(x_i^{r_i}-1,N)$, where $x_i\neq a$ is a random element of $\mathbb Z_N^*$ and $r_i$ is usually a multiple of the largest odd divisor of $r$ (see below). We remark that conceptually, it does not matter whether the green section is performed by a quantum computer simulator or a real quantum computing device.}
  \label{fig:scheme}
\end{figure*}

To study the performance of Shor's algorithm for much larger integers than what can be tested on real quantum devices, we have developed a massively parallel simulator called \texttt{shorgpu}~\cite{shorgpu}, specifically designed to execute the iterative Shor algorithm on multiple GPUs.
Using \texttt{shorgpu}, we have examined over 60000 factoring scenarios for integers up to $N_{\mathrm{max}}=549755813701$, significantly surpassing previous achievements using statevector simulators~\cite{DeRaedt2007MassivelyParallel,DeRaedt2018MassivelyParallel,tankasala2020quantumkit}, matrix product states~\cite{Wang2017ShorSimulationMatrixProductStates1,Dang2019ShorSimulationMatrixProductStates2} (in~\cite{Dang2019ShorSimulationMatrixProductStates2}, the authors simulated 60 qubits to factor $N=961307$), and tensor networks~\cite{Dumitrescu2017TensorNetworShor,Zhao2021ShorTensorNetworkOnSupercomputers}. Note that $N_{\mathrm{max}}$ is still ``small'' for cryptographic purposes.
In order to handle integers of the size of $N_{\mathrm{max}}$, \texttt{shorgpu} uses a new technique (see Section~\ref{sec:2}) to perform the distributed memory communications.

To factor $N_{\mathrm{max}}$, the conventional Shor algorithm would need 117 qubits. The iterative Shor algorithm, however, needs only 40 qubits. It is important to note that the resulting quantum algorithm is still an honest implementation of Shor's algorithm: It produces the same results, does not require exponentially large classical resources (given a large enough quantum computer) and, most importantly, does not exploit a-priori knowledge of the factors~\cite{Smolin2013OversimplifyingShorFactoring}.

We emphasize that for all our simulations, we do not require the solution of the factoring problem to be known. If one presumes knowledge of the solution, and one is not interested in simulating the effect of quantum errors, it is possible to study even larger, cryptographically relevant cases using Qunundrum~\cite{Ekera2020Qunundrum}.

The procedure used to factor integers is shown in Fig.~\ref{fig:scheme}: First, a factoring problem is selected, consisting of a semiprime $N=p\times q$ to factor and a random integer $1<a<N$ comprime to $N$ (i.e., $\mathrm{gcd}(a,N)=1$). Using this as input, \texttt{shorgpu} executes the iterative Shor algorithm with $n=L+1$ qubits to produce several bitstrings. Each bitstring $j$ is processed using either Shor's~\cite{shor1994factoring,ekert1996quantumalgorithms,shor1997algorithm,NielsenChuang} or Eker\aa's~\cite{Ekera2021OnCompletelyFactoringAnyIntegerShor,Ekera2022OnTheSuccessProbabilityOfQuantumOrderFindingShor} post-processing method, which may or may not produce a factor of $N$ (see the yellow section in Fig.~\ref{fig:scheme}). An important step on the way is to extract a candidate $r$ for the \emph{order} $\hat r=\mathrm{ord}_N(a)$. Here $\mathrm{ord}_N(a)$ denotes the order of $a$ modulo $N$, defined as the smallest exponent $\hat r>0$ such that $a^{\hat r}\,\mathrm{mod}\,N=1$. 

Note that ``success'' is not guaranteed by Shor's algorithm; in particular, the sampled bitstring might produce an $r\neq\hat r$ that is not the order, or $r$ might be odd, in which case Shor's post-processing method is not guaranteed to work. However, if the blind application of Shor's factoring procedure still yields at least one factor, we count this execution as a ``lucky'' case. As shown below, a ``lucky'' factor is found much more often than expected.

In principle, the green section in Fig.~\ref{fig:scheme} representing \texttt{shorgpu} can be completely replaced by a sufficiently large, error-corrected quantum computing device. With this in mind, we put forward the challenge of \emph{indirect quantum supremacy}~\cite{Google2019QuantumSupremacy} (a.k.a.~\emph{limited quantum speedup}~\cite{Ronnow2014DefiningQuantumSpeedup}) for a future quantum computer. Here, ``indirect'' means that the simulator (running on a conventional computer) is required to simulate an ideal quantum computational model that executes the same quantum algorithm as the quantum computer, without using any prior knowledge of the solution. More specifically, the challenge for a gate-based quantum computer would be to factor, using Shor's algorithm without oversimplification~\cite{Smolin2013OversimplifyingShorFactoring}, an ``interesting'' semiprime---where ``interesting'' means that the two distinct prime factors shall have the same number of digits---that is larger than the largest semiprime that can be factored by the quantum computer simulator.

\subsection{Related work}
\label{sec:relatedwork}

There is a large body of literature on Shor's quantum factoring algorithm. They can be roughly classified into five main categories. In this section, we give a survey of their main goals and discuss several individual results.

\begin{enumerate}
    \item \textbf{Theory:} The first class of articles focuses on theoretical perspectives such as algorithmic modifications and improved lower bounds on the success probability~\cite{Knill1995ShorIncreasingProbabilityTradeoff,DiVincenzo1995QuantumComputation,Barenco1996ApproximateQFT,Vedral1996QuantumNetworksElementaryArithmeticShor,Parker2000EfficientShorSingleQubitAndLogNMixedQubits,Seifert2001FewerQubitsShorsAlgorithm,McAnally2001RefinementOfShorsAlgorithm,Leander2002ImprovingSuccessProbabilityShor,Coppersmith2002ApproximateQFTForShor,Beauregard2003ShorWith2nplus3Qubits,Fowler2004ShorWithImpreciseRotationGates,Kendon2004EntanglementAndItsRoleInShorsAlgorithm,Gerjuoy2005ShorProbabilityImprovement,VanMeter2005FastQuantumModularExponentiationShor,Devitt2006RobustnessOfShorsAlgorithm,Zalka2006ShorWithFewerQubits,Bourdon2007SharProbabilityEstimatesShor,Markov2012ShorOptimizedModularMultiplication,Markov2013ShorFasterCircuitSynthesis,Grosshans2015FactoringSafeSemiprimesShor,Lawson2015OddOrdersShor,Johnston2017OddOrdersShor,Haner2017ShorWith2nplus2QubitsToffoli,Davis2021ShorForBenchmarking,Bastos2021DetectWhetherShorSucceeds,Antipov2022QuantumPrimitivesForShorsAlgorithm}, many of which consider the case that some parameters of Shor's algorithm are modified and certain trade offs are made. To this class belongs work that estimates the number of resources required when using different levels of quantum computer technology~\cite{Gidney2021HowToFactor2048RSAShor,Gouzien2021ShorFactoring2048RSAIn177DaysWith13436Qubits}. This line of work culminates in {Eker\aa}'s post-processing algorithms~\cite{Ekera2022OnTheSuccessProbabilityOfQuantumOrderFindingShor}, by which the success probability for a single run of the quantum part can be brought arbitrarily close to one (see below).

    \item \textbf{Simulation:} Second, Shor's algorithm has been studied by using simulators running on conventional computers. Some use universal quantum computer simulators~\cite{DeRaedt2007MassivelyParallel,Nam2012ShorPerformanceBandedQFT,DeRaedt2018MassivelyParallel,tankasala2020quantumkit}, sometimes also called Schr\"odinger simulators since they propagate the full quantum statevector. Another approach is to use so-called Feynman simulators, which can only access certain amplitudes from the full statevector, but may require less computational resources; they are often based on tensor networks or matrix product states ~\cite{Dumitrescu2017TensorNetworShor,Wang2017ShorSimulationMatrixProductStates1,Dang2019ShorSimulationMatrixProductStates2,Zhao2021ShorTensorNetworkOnSupercomputers}. Finally, there is software designed to directly sample from the probability distributions generated by Shor's algorithm (cf.~Eq.~(\ref{eq:distribution}) below) and various extensions thereof. To this class belongs the suite of programs called Qunundrum~\cite{Ekera2021QuantumAlgorithmsWithTradeoffsSamplingShor,Ekera2020Qunundrum}, which can simulate distributions for large, cryptographically relevant cases. Note, however, that the solution to the factoring problem (i.e., the order or the discrete logarithm) must be known in advance, and the effect of errors in the quantum part cannot be simulated.

    \item \textbf{Alternative Algorithms:} A third line of work studies alternative ways to use gate-based quantum computers to solve the factoring problem. Some of them use Shor's discrete logarithm quantum algorithm~\cite{Ekera2017FactoringWithDiscreteLogarithm,Ekera2020OnPostProcessingInShor,Ekera2021QuantumAlgorithmsWithTradeoffsSamplingShor}, which is also an instance of the hidden subgroup problem~\cite{Jozsa2001hiddensubgroupproblem}. In the Eker\aa-H\aa{}stad scheme~\cite{Ekera2017FactoringWithDiscreteLogarithm}, the idea to factor a semiprime $N=pq$ is to pick a random $g\in\mathbb Z_N^*$, compute $y=g^{N+1}\,\mathrm{mod}\,N$ with unknown order $r$, and then obtain $d\equiv\log_g{y}\equiv pq+1\equiv pq+1-\phi(N)\equiv p+q\ (\mathrm{mod}\,r)$ (using that $r\mid\phi(N)=(p-1)(q-1)$~\cite{HardyWright}). If $r>p+q$ (which is the case for many $g$), we have $d=p+q$, and additionally knowing $N=pq$ allows one to compute $p$ and $q$. Another alternative way to solve the factoring problem is given in~\cite{Bernstein2017GroverAcceleratedQuantumFactoring} and is based on the classical number field sieve~\cite{lenstra1993developmentOfTheNumberFieldSieve}. In particular, Bernstein et al.~propose to use Grover's quantum search algorithm~\cite{Grover1996Algorithm} (and/or Shor's algorithm for a much smaller subproblem) to accelerate the number field sieve. This proposal is asymptotically worse in time complexity than using Shor's algorithm directly, but it needs less qubits and is therefore possibly easier to realize in near-term physical devices.
    Finally, Li et.~al.~\cite{Li2012QuantumAlgorithmSquareFreeDecomposition} presented an algorithm with an exponential speedup (beyond the framework of the hidden subgroup problem~\cite{Jozsa2001hiddensubgroupproblem}) that solves the square-free decomposition problem---a problem related to factoring in which the task is to find, for any integer $N>0$, the unique integers $N_r$ and $N_s^2$ of the square-free decomposition $N=N_rN_s^2$.

    \item \textbf{Gate-Based Experiments:} Fourth, there have been several experimental efforts to implement Shor's factoring algorithm on existing gate-based quantum computer devices~\cite{Vandersypen2001ExperimentalShorNMR,Lu2007CompiledShorPhotonicQubits,Lanyon2007ExperimentalCompiledShorPhotons,Politi2009ShorPhotonicChip,Lucero2012ShorJosephsonPhaseQubits,MartinLopez2012ShorExperimentQubitRecycling,Monz2016ShorOnTrappedIons,Amico2019ShorOnIBMQ,Skosana2021ShorFactoring21IBM,Abhijith2022QuantumAlgorithmImplementationsForBeginners}. However, many of these have made use of prior knowledge about the factors to simplify the experimental setup~\cite{Smolin2013OversimplifyingShorFactoring}. In the extreme case (namely when a base $a\in\mathbb Z_N^*$ with order $\mathrm{ord}_N(a) = 2$ is used), this brings the computational problem down to flipping coins. Experiments that have not used such a kind of oversimplification can be found in~\cite{MartinLopez2012ShorExperimentQubitRecycling,Monz2016ShorOnTrappedIons,Amico2019ShorOnIBMQ}.

    \item \textbf{Other Experiments:} Finally, quantum annealers and adiabatic quantum computers have been used to study alternative factoring algorithms~\cite{Peng2008QuantumAdiabaticAlgorithmForFactorization,Andriyash2016BoostingIntegerFactorizationDWave,Dridi2017PrimeFactorizationDWave,Jiang2018QuantumAnnealingForPrimeFactorization,Peng2019FactoringLargeIntegersDWave,Mengoni2020BreakingRSAWithDWave2000Q,Wang2020PrimeFactorizationParemeterOptimizationIsingAnnealer}. The quantum annealing approach requires at most $O(L^2)$ qubits to factor an $L$-bit number. Quantum annealing and adiabatic quantum computation are technologically significantly ahead of gate-based quantum computing, in that larger quantum processing units with more than 5000 qubits exist and that they can solve much larger problems~\cite{King2022CoherentQuantumAnnealing2000IsingChain,King2023QuantumCriticalDynamics5000SpinGlass}. In particular, numbers up to and above 200000 have been factored on the D-Wave 2000Q~\cite{Dridi2017PrimeFactorizationDWave,Jiang2018QuantumAnnealingForPrimeFactorization} and 1005973 has been factored using D-Wave hybrid~\cite{Peng2019FactoringLargeIntegersDWave}. Although significantly larger than the numbers factored on gate-based quantum computers (without oversimplification), these numbers are still much smaller than $N_{\mathrm{max}}=549755813701$ factored in this work using \texttt{shorgpu}.
\end{enumerate}

\subsection{Outline}

This paper is structured as follows. 
In Section~\ref{sec:2}, we describe the algorithmic details of \texttt{shorgpu}. In particular, we explain how to implement the modular multiplication as a systematic communication scheme between the compute nodes. 
In Section~\ref{sec:3}, we present our results from over 60000 quantum computer simulations using up to 2048 GPUs.
Section~\ref{sec:4} contains our conclusions.

\section{Simulation}
\label{sec:2}

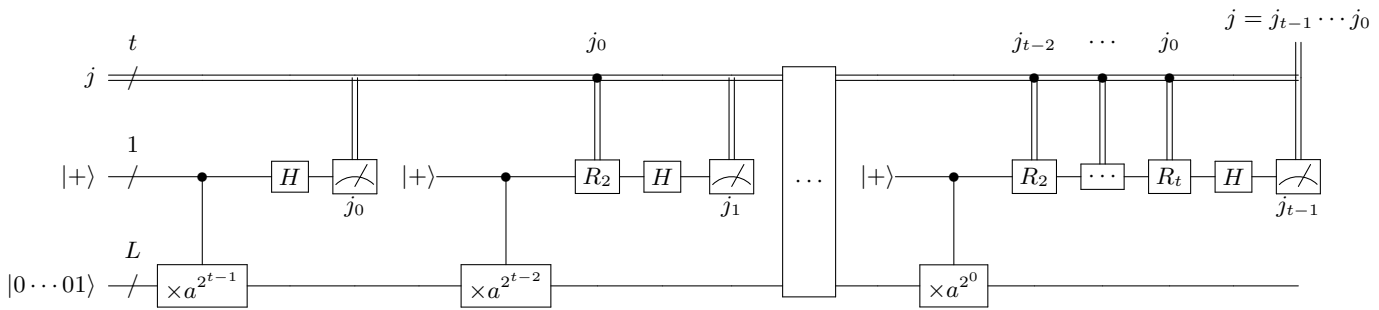
\begin{figure*}
\centering
\begin{equation*}
    \Qcircuit @C=1.0em @R=1.0em {
        & t &&&&&& j_0 &&&&&& j_{t-2} & \cdots & j_0 && \ustick{j=j_{t-1}\cdots j_0} \\
        \lstick{j} & \cw{/} & \cw & \cw & \cw\cwx[2] & \cw & \cw & 
            \cctrl{2} & \cw & \cw\cwx[2] & \cghost{\cdots} & \cw & \cw & \cctrl{2} &
            \cctrl{2} & \cctrl{2} & \cw & \cw\cwx[2]\cwx[-1]\\
        & \dstick{1} &&&&&&&& \nghost{\cdots} \\
        \lstick{\ket{+}} & \qw{/} & \ctrl{2} & \gate{H} & \meter & \push{\ket{+}} & \ctrl{2} &
            \gate{R_2} & \gate{H} & \meter & \nghost{\cdots} & \push{\ket{+}} & \ctrl{2} & \gate{R_2} &
            \gate{\cdots}  & \gate{R_t} & \gate{H} & \meter\\
        & \dstick{L} &&& \ustick{j_0} &&&&& \ustick{j_1} & \nghost{\cdots} &&&&&&& \ustick{j_{t-1}}\\
        \lstick{\ket{0\cdots 01}} & \qw{/} & \gate{\times a^{2^{t-1}}} & \qw & \qw & \qw & \gate{\times a^{2^{t-2}}} &
            \qw & \qw & \qw& \multigate{-4}{\cdots} & \qw & \gate{\times a^{2^{0}}} & \qw &
            \qw & \qw & \qw & \qw\\
    }
\end{equation*}
\caption{\textbf{Quantum circuit of the iterative Shor algorithm.} The circuit consists of $L+1$ qubits that undergo $t$ separate stages $\texttt{cbit}=0,\ldots,t-1$, in which the classical bit $j_{\texttt{cbit}}$ is measured. Each stage starts with the first qubit in the initial state $\ket+$ and ends with this qubit being measured (middle row). Between initialization and measurement, each stage consists of a controlled modular multiplication (bottom row) with some power of $a$ (see Eq.~(\ref{eq:oraclegate})), then a rotation gate controlled by all previously measured classical bits  (see Eq.~(\ref{eq:phasegate})), and finally a Hadamard gate. The resulting bit $j_{\texttt{cbit}}$ is used to assemble the classical bitstring $j=j_{t-1}\cdots j_0$ (top row).}
\label{fig:shor}
\end{figure*}

For almost all results reported in this work, we use \texttt{shorgpu} to simulate the iterative Shor algorithm (the source code is available online~\cite{shorgpu}). It propagates an $n=L+1$-qubit statevector $\ket\psi$ consisting of $2^{L+1}$ complex numbers through the quantum circuit for factoring an $L$-bit number shown in Fig.~\ref{fig:shor}. Each step in the quantum circuit corresponds to an operation on $\ket\psi$. In this section, we describe each of these operations in a linear algebra context. The probability distribution generated by Shor's algorithm is derived and visualized in Appendix~\ref{app:distribution}.

As the total memory is the bottleneck of such statevector simulations, the $2^{L+1}$ complex numbers $\ket\psi$ are distributed over the memory of up to 2048 GPUs (cf.~Fig.~\ref{fig:scheme}). Communication between the GPUs is managed using the Message Passing Interface (MPI)~\cite{mpi40}. 

We use the J\"ulich Universal Quantum Computer Simulator (JUQCS)~\cite{DeRaedt2007MassivelyParallel,DeRaedt2018MassivelyParallel,Willsch2021JUQCSGQAOA} for verification. JUQCS was previously used to simulate the conventional Shor algorithm for $N\le 65531$ with up to $n=48$ qubits on the Sunway TaihuLight and the K computer~\cite{DeRaedt2018MassivelyParallel}.
A few features had to be added to JUQCS to be able to also simulate the iterative Shor algorithm. The latter made it possible to simulate one bitstring for $N=4194293$ with $n=23$ qubits in about 720 seconds (using 4 A100 GPUs). However, our JUQCS implementation of the oracle which performs the modular exponentiation becomes highly inefficient as the number $n$ of qubits increases because it does not distribute well over many cores or GPUs.
In contrast, the new, dedicated algorithm described below generates a bitstring in about 0.4 seconds for the same problem and the same number of GPUs. For the largest problem simulated ($N=549755813701$ with $n=40$ qubits), \texttt{shorgpu} generates a bitstring in about 200 seconds using 2048 GPUs.
We verified that the iterative Shor algorithm simulated with \texttt{shorgpu} produces the same results as JUQCS for problems of the size that can be simulated with JUQCS.

\subsection{Initialization}
\label{sec:initialization}

To simulate the iterative Shor algorithm for factoring an $L$-bit semiprime $N$, \texttt{shorgpu} simulates the full quantum circuit with $n=L+1$ qubits shown in Fig.~\ref{fig:shor}.
This is done by computing all complex coefficients of the statevector
\begin{align}
    \label{eq:psicoefficients}
    \ket\psi 
             = \sum_{k_L\cdots k_0=0,1} \psi_{k_L\cdots k_0}\ket{k_L\cdots k_0}
             = \begin{pmatrix}
                \psi_{0\cdots00}\\
                \psi_{0\cdots01}\\
                \vdots\\
                \psi_{1\cdots11}
             \end{pmatrix}.
\end{align}
These $2^{L+1}$ complex double-precision numbers $\psi_{k_L\cdots k_0}$ are distributed over $N_{\text{GPU}} \in\{2,4,8,\ldots,2048\}$ GPUs. The distributed memory communication between the GPUs uses CUDA-aware MPI. 
In our approach, each GPU is identified by its MPI rank, i.e., an $n_{\text{global}}$-bit integer called $\texttt{mpi\_rank}=0,\ldots,N_{\text{GPU}}-1$. Here, $n_{\text{global}}$ denotes the number of so-called \emph{global qubits} (see~\cite{DeRaedt2007MassivelyParallel,DeRaedt2018MassivelyParallel,Willsch2021JUQCSGQAOA}). We have $N_{\text{GPU}}=2^{n_{\text{global}}}$ GPUs. The other $n_{\text{local}} = n-n_{\text{global}}$ qubits are called \emph{local qubits}, since each GPU holds in its local memory all $2^{n_{\text{local}}}$ complex coefficients
\begin{align}
   \begin{pmatrix}
        \psi_{\mathrm{bin}(\texttt{mpi\_rank})0\cdots00}\\
        \psi_{\mathrm{bin}(\texttt{mpi\_rank})0\cdots01}\\
        \vdots\\
        \psi_{\mathrm{bin}(\texttt{mpi\_rank})1\cdots11}
     \end{pmatrix}.
\end{align}
The GPUs (i.e., the MPI processes) are further divided into two separate groups, identified by the most significant bit of the MPI rank,
\begin{align}
    \label{eq:mpirank}
    \mathrm{bin}(\texttt{mpi\_rank}) = \texttt{mpi\_x}\:\mathrm{bin}(\texttt{mpi\_xrank}).
\end{align}
Here, $\texttt{mpi\_x}=0,1$ identifies the group and $0\le\texttt{mpi\_xrank}<N_{\text{GPU}}/2$ identifies the GPU within each group. Thus, \texttt{shorgpu} requires at least two GPUs to work (unless a single GPU is used with 2 MPI processes in overscheduling mode). The reason for the separation into two groups is that the implementation of the controlled modular multiplication gate (see below) requires an all-to-all communication between all GPUs with $\texttt{mpi\_x}=1$.

At the start of the simulation, the statevector $\ket\psi$ is initialized in the state $\ket+\ket{0\cdots01}$, where $\ket+=(\ket0+\ket1)/\sqrt2$. This means that we set
\begin{align}
    \psi_{00\cdots01} &= \frac 1 {\sqrt2}, \\
    \psi_{10\cdots01} &= \frac 1 {\sqrt2},
\end{align}
and all other coefficients to zero. This type of initialization is always used unless \texttt{shorgpu} is used to assess the effect of quantum initialization errors (for information on this mode, see Section~\ref{sec:resetoperation} below).

\subsection{Controlled Modular Multiplication Gate}
\label{sec:multiplicationgate}

\begin{figure*}
  \centering
  \includegraphics[width=\textwidth]{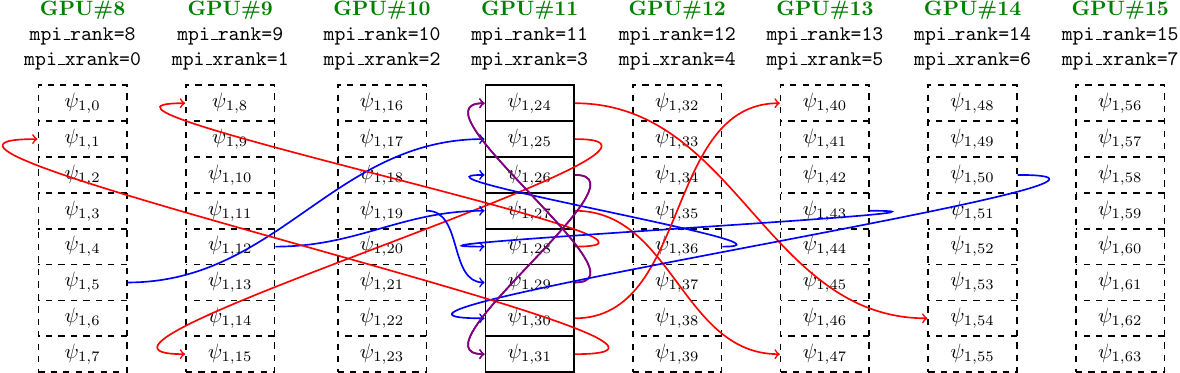}
  \caption{\textbf{Illustration of the MPI communication scheme for the implementation of the controlled modular multiplication gate given by Eq.~(\ref{eq:oraclegate}) for the quantum circuit to factor $N=55=5\times11$ with $a=16$.} This circuit needs $n=7$ qubits, i.e., the first qubit for the measurement (middle line in Fig.~\ref{fig:shor}) and $L=6$ qubits to represent $N$. There are $N_{\mathrm{GPU}}=16$ GPUs in this example, so we have $n_{\mathrm{global}} = 4$ global and $n_{\mathrm{local}} = 3$ local qubits (note that this is only for illustration purposes; in practice one would use much fewer GPUs for 7 qubits). Shown is the implementation of the last oracle gate in Fig.~\ref{fig:shor} (the controlled modular multiplication with $\texttt{a}=a^{2^0}=16$) from the perspective of GPU\#11. $\psi_{x,y}$ denotes, in the notation of Eq.~(\ref{eq:psicoefficients}), the statevector coefficient $\psi_{x\,\mathrm{bin}(y)}$. Red arrows represent coefficients that are sent from GPU\#11 to another GPU (whose index is computed from $\texttt{a}y\,\mathrm{mod}\,N$.) Blue arrows represent coefficients that are sent to GPU\#11 from another GPU (whose index is computed from $\texttt{a}^{-1}y\,\mathrm{mod}\,N$). Purple arrows represent coefficients that stay on GPU\#11. Note that GPU\#15 is not involved in the communication because $N\le y$ for all $\psi_{x,y}$ of GPU\#15, so the oracle gate does not permute these coefficients (the last case in Eq.~(\ref{eq:oraclegate})).
  }
  \label{fig:mpischeme}
\end{figure*}

The first gate in each of the $t$ stages in Fig.~\ref{fig:shor} is the controlled modular multiplication gate (also called \emph{oracle gate}), controlled by the first qubit. Mathematically, its operation is defined by
\begin{align}
    \label{eq:oraclegate}
    \mathrm{C}U_{\texttt{a}}\ket x\ket y = \begin{cases}
        \ket{0}\ket{y} & (x=0) \\
        \ket{1}\ket{\texttt{a}y\,\mathrm{mod}\,N} & (x=1\text{ and }0\le y<N) \\
        \ket{1}\ket{y} & (x=1\text{ and }N\le y)
    \end{cases},
\end{align}
where $x$ denotes the first qubit, $y=0,\ldots,N_{\text{GPU}}/2$ denotes the other qubits, and $\texttt{a}\in\{a^{2^{t-1}}\,\mathrm{mod}\,N, a^{2^{t-2}}\,\mathrm{mod}\,N, \ldots, a\}$ stands for one of the powers of $a$ in Fig.~\ref{fig:shor}. Note that each individual modular exponentiation is always precomputed for \emph{any} realization of this circuit (we use the shift-and-multiply algorithm). This is independent of whether the circuit is executed by a quantum computer simulator or a real quantum computer (see also \cite{Monz2016ShorOnTrappedIons,Amico2019ShorOnIBMQ,Gidney2021HowToFactor2048RSAShor}).

Looking at Eq.~(\ref{eq:oraclegate}), we see that the oracle gate performs a permutation of all complex coefficients among the GPUs in the $\texttt{mpi\_x}=1$ group.
\texttt{shorgpu} implements this unitary operation by computing, on each GPU, all indices of the coefficients that are sent to other GPUs (stored in a GPU buffer \texttt{oracle\_idxsend}) and those that are received from other GPUs (stored in a GPU buffer \texttt{oracle\_idxrecv}) using the precomputed modular inverse $\texttt{ainv} = \texttt{a}^{-1}\,\mathrm{mod}\,N$, which is efficiently computable using the extended Euclidean algorithm. The MPI communication scheme for an example with 16 GPUs is shown in Fig.~\ref{fig:mpischeme}.

The complexity of the permutation depends on the value of $a$. For instance, in the special case that the order of $a$ is a small power of 2, we have $\texttt{a}=1$ in many of the early stages, so the oracle gate would not require MPI communication between different GPUs. In general the communication scheme can be very complicated. Figure~\ref{fig:mpischeme} shows a typical instance where each GPU sends (red arrows) and receives (blue arrows) some coefficients from other GPUs.

To implement this communication scheme between the GPUs, \texttt{shorgpu} uses non-blocking point-to-point communication in a circular fashion using \texttt{MPI\_Isend} and \texttt{MPI\_Irecv}. Additionally, before the send operations, each GPU first arranges all coefficients that are sent to a particular other GPU in a contiguous block of memory, schematically denoted by $\psi^{(\mathrm{cont})}$. This is imperative since for large $N$, this part of the simulation takes a significant fraction of the total run time. Alternative implementations using one-sided communication such as \texttt{MPI\_Put}, collective communication using \texttt{MPI\_Alltoallv}, or communication based on custom MPI data types (see~\cite{mpi40} for more information) performed significantly worse in our experiments.

\subsection{Rotation Gate}

After the oracle gate, each stage (except the first stage) of the quantum circuit in Fig.~\ref{fig:shor} contains a sequence of rotation gates defined by
\begin{align}
    R_l = \begin{pmatrix}
         1 & 0 \\
         0 & e^{2\pi i/2^l}
    \end{pmatrix}.
\end{align}
These are controlled by bits resulting from previous measurements. Specifically, at the stage $\texttt{cbit}=0,\ldots,t-1$, in which the classical bit $j_{\texttt{cbit}}$ is being measured, the sequence of these controlled rotation gates reads
\begin{align}
    \label{eq:phasegate}
    \prod_{l=2}^{1+\texttt{cbit}} \mathrm{C}R_l 
    = \prod_{l=2}^{1+\texttt{cbit}} \begin{pmatrix}
         1 & 0 \\
         0 & e^{2\pi i j_{1+\texttt{cbit}-l}/2^l}
    \end{pmatrix}
    = \begin{pmatrix}
         1 & 0 \\
         0 & e^{i\varphi_{\texttt{cbit}}}
    \end{pmatrix},
\end{align}
where the phase $\varphi_{\texttt{cbit}}$ at stage $\texttt{cbit}$ amounts to
\begin{align}
    \varphi_{\texttt{cbit}} = 2\pi\sum_{l=2}^{1+\texttt{cbit}} \frac{j_{1+\texttt{cbit}-l}}{2^l} = \frac{\pi j^{(\texttt{cbit})}}{2^{\texttt{cbit}}}, 
\end{align}
and $j^{(\texttt{cbit})} = j_{\texttt{cbit}-1}j_{\texttt{cbit}-2}\cdots j_{1}j_0$ is the integer assembled from all classical bits measured up to this point.

As the phase gate given by Eq.~(\ref{eq:phasegate}) only affects coefficients $\psi_{k_L\cdots k_0}$ where the first qubit index $k_L=1$, this operation only needs to be implemented by the GPUs in the $\texttt{mpi\_x}=1$ group. This is done directly after the implementation of the oracle gate, when moving all coefficients out of the contiguous memory blocks $\psi^{(\mathrm{cont})}$, according to
\begin{align}
    \mathrm{Re}(\psi_{1\,\mathrm{bin}(\texttt{mpi\_xrank})\,*\cdots*}&)\nonumber\\
        \leftarrow \cos(\varphi_{\texttt{cbit}})& \mathrm{Re}(\psi_{1\,\mathrm{bin}(\texttt{mpi\_xrank})\,*\cdots*}^{(\mathrm{cont})})\nonumber\\
        - \sin(\varphi_{\texttt{cbit}})& \mathrm{Im}(\psi_{1\,\mathrm{bin}(\texttt{mpi\_xrank})\,*\cdots*}^{(\mathrm{cont})}), \\
    \mathrm{Im}(\psi_{1\,\mathrm{bin}(\texttt{mpi\_xrank})\,*\cdots*}&)\nonumber\\
        \leftarrow \cos(\varphi_{\texttt{cbit}})& \mathrm{Im}(\psi_{1\,\mathrm{bin}(\texttt{mpi\_xrank})\,*\cdots*}^{(\mathrm{cont})})\nonumber\\
        + \sin(\varphi_{\texttt{cbit}})& \mathrm{Re}(\psi_{1\,\mathrm{bin}(\texttt{mpi\_xrank})\,*\cdots*}^{(\mathrm{cont})})    .
\end{align}

\subsection{Hadamard Gate}

The implementation of the Hadamard gate on the first qubit transforms the statevector coefficients as
\begin{align}
    &\psi_{0\,\mathrm{bin}(\texttt{mpi\_xrank})\,*\cdots*}\nonumber\\
    &\quad\leftarrow \frac{\psi_{0\,\mathrm{bin}(\texttt{mpi\_xrank})\,*\cdots*} + \psi_{1\,\mathrm{bin}(\texttt{mpi\_xrank})\,*\cdots*}}{\sqrt2}, \\
    &\psi_{1\,\mathrm{bin}(\texttt{mpi\_xrank})\,*\cdots*}\nonumber\\
    &\quad\leftarrow \frac{\psi_{0\,\mathrm{bin}(\texttt{mpi\_xrank})\,*\cdots*} - \psi_{1\,\mathrm{bin}(\texttt{mpi\_xrank})\,*\cdots*}}{\sqrt2}.
\end{align}
For every GPU in the $\texttt{mpi\_x}=0$ group, this requires two-sided MPI communication with exactly one GPU in the  $\texttt{mpi\_x}=1$ group.

\subsection{Measurement Operation}

At the end of each stage in Fig.~\ref{fig:shor}, the classical bit $j_{\texttt{cbit}}$ is measured, where $\texttt{cbit}=0,\ldots,t-1$ enumerates the stage. This amounts to adding up the probabilities 
\begin{align}
    \label{eq:probabilityp1measurement}
    p_1 = \sum_{k_{L-1}\cdots k_0=0,1} \vert \psi_{1\,k_{L-1}\cdots k_0} \vert^2,
\end{align}
which is an MPI reduction over all GPUs belonging to the $\texttt{mpi\_x}=1$ group. The probability to measure 1 (0) is then given by $p_1$ ($p_0=1-p_1$). This probability can be used to sample $j_{\texttt{cbit}}$, which is done by drawing a uniform random number $R\in[0,1)$, and assigning $j_{\texttt{cbit}}=1$ if this $R<p_1$ and $j_{\texttt{cbit}}=0$ otherwise.

\subsection{Reset Operation}
\label{sec:resetoperation}

The reset operation performs both the von-Neumann projection of the statevector to the result of the measurement and the reinitialization of the first qubit in $\ket+$ at the same time. If the result of the measurement is given by $j_{\texttt{cbit}}=0,1$, this operation is done by transforming all coefficients according to
\begin{align}
    \begin{pmatrix}
        \psi_{00\cdots0}\\
        \vdots \\
        \psi_{01\cdots1}\\
        \psi_{10\cdots0}\\
        \vdots \\
        \psi_{11\cdots1}\\
     \end{pmatrix} 
    \leftarrow
    \begin{pmatrix}
        \psi_{j_{\texttt{cbit}}0\cdots0}/\sqrt{2p_{j_{\texttt{cbit}}}}\\
        \vdots \\
        \psi_{j_{\texttt{cbit}}1\cdots1}/\sqrt{2p_{j_{\texttt{cbit}}}}\\
        \psi_{j_{\texttt{cbit}}0\cdots0}/\sqrt{2p_{j_{\texttt{cbit}}}}\\
        \vdots \\
        \psi_{j_{\texttt{cbit}}1\cdots1}/\sqrt{2p_{j_{\texttt{cbit}}}}\\
     \end{pmatrix}.
\end{align}
Of course, in the case of quantum errors, the coefficients have to be replaced accordingly (cf.~Eqs.~(\ref{eq:quantumerrorstatecorrect}) and (\ref{eq:quantumerrorstateerror})).

This operation requires an MPI transfer of all coefficients from the GPUs in the group $\texttt{mpi\_x}=j_{\texttt{cbit}}$ to the GPUs in the group $\texttt{mpi\_x}=1-j_{\texttt{cbit}}$.

\subsection{Initialization Errors}

There are two different types of initialization errors that \texttt{shorgpu} can simulate, namely an amplitude initialization error and a phase initialization error. In both cases, a slightly different initial state $\ket{+'}$ is used instead of $\ket{+}$ for the first qubit in all stages $\texttt{cbit}=0,\ldots,t-1$ of the circuit in Fig.~\ref{fig:shor}. The slightly erroneous state $\ket{+'}$ is parameterized in terms of an error parameter $\delta\in[0,1]$. Our motivation to prioritize the recycled qubit for a study of initialization errors instead of the other ``internal'' qubits is that this qubit is measured and reinitialized successively in every stage of the iterative Shor algorithm.

\subsubsection{Amplitude Initialization Error}
\label{sec:amplitudeerror}

We define an amplitude initialization error as the case in which, at the beginning of each stage in Fig.~\ref{fig:shor}, the quantum state is not initialized in the equal superposition $\ket+$ but the slightly unequal superposition
\begin{align}
    \ket{+_{\mathrm{ampl}}'(\delta)} = \sqrt{\frac{1+\delta}2}\ket0 + \sqrt{\frac{1-\delta}2}\ket1.
\end{align}
This expression is motivated by the observation that quantum computer prototypes from the NISQ era sometimes tend to prefer $\ket0$ over $\ket1$ when brought to a uniform superposition by multiple quantum gates~\cite{Michielsen2017BenchmarkingQC}. Furthermore, one of the most prominent decoherence and noise processes in qubit systems is a decay from $\ket1$ to $\ket0$, a so-called $T_1$ relaxation process~\cite{Weiss2012QuantumDissipativeSystems,Paladino2015OneOverFNoiseRelaxationTime,Carroll2022DynamicsSuperconductingQubitRelaxationTimes}.

\subsubsection{Phase Initialization Error}
\label{sec:phaseerror}

As a second type of initialization error, we consider a phase initialization error defined as
\begin{align}
    \ket{+_{\mathrm{phase}}'(\delta)} = \frac{1}{\sqrt{2}}\ket0 + \frac{e^{i\pi\delta}}{\sqrt{2}}\ket1.
\end{align}
This expression is motivated by the fact that besides $T_1$ relaxation, dephasing processes are another prominent consequence of decoherence and noise in quantum systems~\cite{fox2006quantumoptics,Weiss2012QuantumDissipativeSystems,Paladino2015OneOverFNoiseRelaxationTime}.

\subsubsection{Effective Single-Qubit Error Probability}

For both initialization errors, the error parameter $\delta\in[0,1]$ can be related to an effective, single-qubit error probability, defined as the probability that the erroneous state would correctly be observed as a $\ket+$ state when measured along the $x$ axis:
\begin{align}
    p_{\mathrm{ampl}}^{\mathrm{error}}(\delta) &= 1 - \vert\braket{+|+_{\mathrm{ampl}}'(\delta)}\vert^2 = \frac{1-\sqrt{1-\delta^2}}2,\\
    p_{\mathrm{phase}}^{\mathrm{error}}(\delta) &= 1 - \vert\braket{+|+_{\mathrm{phase}}'(\delta)}\vert^2 = \frac{1-\cos(\pi\delta)}2.
\end{align}
Note, however, that this interpretation is not unique; depending on the particular realization of the quantum circuit, there may be more reasonable, alternative interpretations of $\delta$ in relation to an effective error probability.

\subsection{Measurement Errors}

For quantum processors, a measurement is often a slow and susceptible process by which destructive influences from the environment can enter the quantum system~\cite{Wallraff2005DispersiveReadout,Gambetta2006CircuitQEDMeasurement,Reed2010HighFidelityReadoutCQED,Jacobs2014QuantumMeasurementTheory,Naghiloo2019QuantumMeasurementSuperconductingQubits}. Moreover, it is particularly challenging to implement quantum non-demolition readout required for midcircuit measurements~\cite{Boissonneault2010ImprovedSuperconductingQubitReadout,corcoles2021exploitingDynamicQuantumCircuits}. We distinguish between two different types of measurement errors, namely a classical error corresponding to a misclassification of the quantum measurement result, and a quantum error that may occur during or before each measurement.

\subsubsection{Classical Measurement Error}
\label{sec:classicalerror}

We define a classical measurement error as a misclassification that occurs right after the quantum measurement process with a given, constant error probability $\delta$. It is defined by flipping only the resulting bit $j_{\texttt{cbit}}$, while leaving the internal quantum state unchanged.

Simulating a classical measurement error requires a second sampling step, by drawing another uniform random number $R_2\in[0,1)$ and flipping the bit if $R_2 < \delta$. In case of a misclassification error, we simply use $j_{\texttt{cbit}}^{(\mathrm{observed})} = 1 - j_{\texttt{cbit}}$ for the classical bitstring in Fig.~\ref{fig:shor}. The quantum state, however, is left in its original state with the first qubit projected on $\ket{j_{\texttt{cbit}}}$.

Note that even such a single misclassification error can have non-trivial consequences, since this error affects the angles of all subsequent rotation gates (see Fig.~\ref{fig:shor}). This has an influence on the measurements of the following bits $\texttt{cbit}+1, \ldots, t-1$. Therefore, a single bit flip error can induce a change in more than one classical bit of the output bitstring $j$.

\subsubsection{Quantum Measurement Error}
\label{sec:quantumerror}

Quantum errors are conventionally modeled as operations $\rho\mapsto\mathcal E(\rho)$ on the system's density matrix $\rho=\ket\psi\!\bra\psi$. If such an operation is a completely positive, trace-preserving map, it is called a \emph{quantum channel} or \emph{error channel} (see \cite{NielsenChuang,Holevo2019QuantumSystemsChannelsInformation,Wilde2017QuantumInformationTheory} for more information).

We model a quantum measurement error by applying a depolarizing error channel in every measurement process (which, on quantum computer hardware, is a time evolution that can take a significant amount of time~\cite{Naghiloo2019QuantumMeasurementSuperconductingQubits}). The depolarizing error channel is defined by the quantum operation
\begin{align}
    \label{eq:depolarizingchannel}
    \mathcal E_{\mathrm{dep}}(\tilde\rho) &= (1-p_x-p_y-p_z)\rho \nonumber\\ 
    &+ p_x\sigma^x\tilde\rho\sigma^x + p_y\sigma^y\tilde\rho\sigma^y + p_z\sigma^z\tilde\rho\sigma^z,
\end{align}
where $\tilde\rho$ is a single-qubit density matrix, $(\sigma^x,\sigma^y,\sigma^z)$ are the Pauli matrices, and $(p_x,p_y,p_z)$ represent the error probabilities for the respective Pauli errors. 
After the application of $\mathcal E_{\mathrm{dep}}$ to the first qubit, the density matrix that describes the state of the full quantum computer reads
\begin{align}
    \rho &= (\mathcal E_{\mathrm{dep}}\otimes I)(\ket\psi\!\bra\psi) \nonumber\\
    &= \sum_{\bar k\bar k'} \mathcal E_{\mathrm{dep}}\Big(
        \psi_{0\bar k}\psi_{0\bar k'}^* \ket0\!\bra0
        + \psi_{0\bar k}\psi_{1\bar k'}^* \ket0\!\bra1 \nonumber\\
        &\qquad+ \psi_{1\bar k}\psi_{0\bar k'}^* \ket1\!\bra0
        + \psi_{1\bar k}\psi_{1\bar k'}^* \ket1\!\bra1 \Big)
        \otimes\ket{\bar k}\!\bra{\bar k'}
        ,
\end{align}
where $I$ is an identity operation on the remaining $L$ qubits, and $\bar k,\bar k'=k_{L-1}\cdots k_0$ enumerate their $2^L$ different indices. 

A measurement of the first qubit is quantum-mechanically described by the measurement operators $\mathcal M_0=\ket0\!\bra0\otimes I$ and  $\mathcal M_1=\ket1\!\bra1\otimes I$. Using Eq.~(\ref{eq:depolarizingchannel}), we find the probability to measure $j_{\texttt{cbit}}=0,1$ as
\begin{align}
    \label{eq:quantumerrormeasurementprobability}
    p_{j_{\texttt{cbit}}}' &= \mathrm{Tr}\,\mathcal M_{j_{\texttt{cbit}}}\rho\mathcal M_{j_{\texttt{cbit}}}^\dagger \nonumber\\
    &= (1-p_x-p_y) p_{j_{\texttt{cbit}}} + (p_x+p_y) p_{1-j_{\texttt{cbit}}},
\end{align}
where $p_{j_{\texttt{cbit}}}=\sum_{\bar k}\vert\psi_{j_{\texttt{cbit}}\bar k}\vert^2$ is computed according to Eq.~(\ref{eq:probabilityp1measurement}). A calculation of the post-measurement state $\rho_{j_{\texttt{cbit}}}'$ yields
\begin{align}
    \rho_{j_{\texttt{cbit}}}'  
    &= \frac{M_{j_{\texttt{cbit}}}\rho\mathcal M_{j_{\texttt{cbit}}}^\dagger}{\mathrm{Tr}\,M_{j_{\texttt{cbit}}}\rho\mathcal M_{j_{\texttt{cbit}}}^\dagger} \nonumber\\
    &= p_{j_{\texttt{cbit}}}^{(\mathrm{correct})} \ket{\psi_{j_{\texttt{cbit}}}^{(\mathrm{correct})}}\!\bra{\psi_{j_{\texttt{cbit}}}^{(\mathrm{correct})}} \nonumber\\
    &+ p_{j_{\texttt{cbit}}}^{(\mathrm{error})} \ket{\psi_{j_{\texttt{cbit}}}^{(\mathrm{error})}}\!\bra{\psi_{j_{\texttt{cbit}}}^{(\mathrm{error})}},
\end{align}
where
\begin{subequations}
    \begin{align}
        \label{eq:quantumerrorstatecorrectprobability}
        p_{j_{\texttt{cbit}}}^{(\mathrm{correct})} 
        &= \frac{(1-p_x-p_y) p_{j_{\texttt{cbit}}}}{(1-p_x-p_y) p_{j_{\texttt{cbit}}} + (p_x+p_y) p_{1-j_{\texttt{cbit}}}},
        \\
        \label{eq:quantumerrorstateerrorprobability}
        p_{j_{\texttt{cbit}}}^{(\mathrm{error})} 
        &= \frac{(p_x+p_y) p_{1-j_{\texttt{cbit}}}}{(1-p_x-p_y) p_{j_{\texttt{cbit}}} + (p_x+p_y) p_{1-j_{\texttt{cbit}}}},
    \end{align}
\end{subequations}
and
\begin{subequations}
    \begin{align}
        \label{eq:quantumerrorstatecorrect}
        \ket{\psi_{j_{\texttt{cbit}}}^{(\mathrm{correct})}} 
        &= \sum_{\bar k} \frac{\psi_{j_{\texttt{cbit}}\bar k}}{\sqrt{p_{j_{\texttt{cbit}}}}} \ket{j_{\texttt{cbit}}\bar k},
        \\
        \label{eq:quantumerrorstateerror}
        \ket{\psi_{j_{\texttt{cbit}}}^{(\mathrm{error})}} 
        &= \sum_{\bar k} \frac{\psi_{(1-j_{\texttt{cbit}})\bar k}}{\sqrt{p_{1-j_{\texttt{cbit}}}}} \ket{j_{\texttt{cbit}}\bar k}. 
    \end{align}
\end{subequations}
Here, the superscript ``correct'' (``error'') refers to the probability and the state in the case that no error (an error) has occurred. Furthermore, the expressions show that both Pauli $x$ and $y$  errors only occur in combination, so we define the joint quantum error probability $\delta = p_x+p_y$, by analogy with the classical case.

As in the classical case, a simulation of the quantum error process requires two sampling operations: First, a random number $R\in[0,1)$ is sampled to assign the measurement result with probability $p_{j_{\texttt{cbit}}}'$ given by Eq.~(\ref{eq:quantumerrormeasurementprobability}), i.e., we assign $j_{\texttt{cbit}}=1$ if this $R<p_1'$ and $j_{\texttt{cbit}}=0$ otherwise. 

Second, a random number $R_2\in[0,1)$ is sampled to determine whether an error has happened or not. If $R_2<\smash{p_{j_{\texttt{cbit}}}^{(\mathrm{error})}}$, an error has happened while measuring $j_{\texttt{cbit}}$, and the simulation continues with the state $\smash{\ket{\psi_{j_{\texttt{cbit}}}^{(\mathrm{error})}}}$ given by Eq.~(\ref{eq:quantumerrorstateerror}). Otherwise, the simulation continues with the state $\smash{\ket{\psi_{j_{\texttt{cbit}}}^{(\mathrm{correct})}}}$.

Note that the quantum error has a more direct influence on the quantum state than the classical error, since the projection to either $\smash{\ket{\psi_{j_{\texttt{cbit}}}^{(\mathrm{correct})}}}$ or $\smash{\ket{\psi_{j_{\texttt{cbit}}}^{(\mathrm{error})}}}$ directly affects the quantum state, not only implicitly through the angles of subsequent rotation gates.

\begin{figure*}
  \centering
  \includegraphics[width=\textwidth]{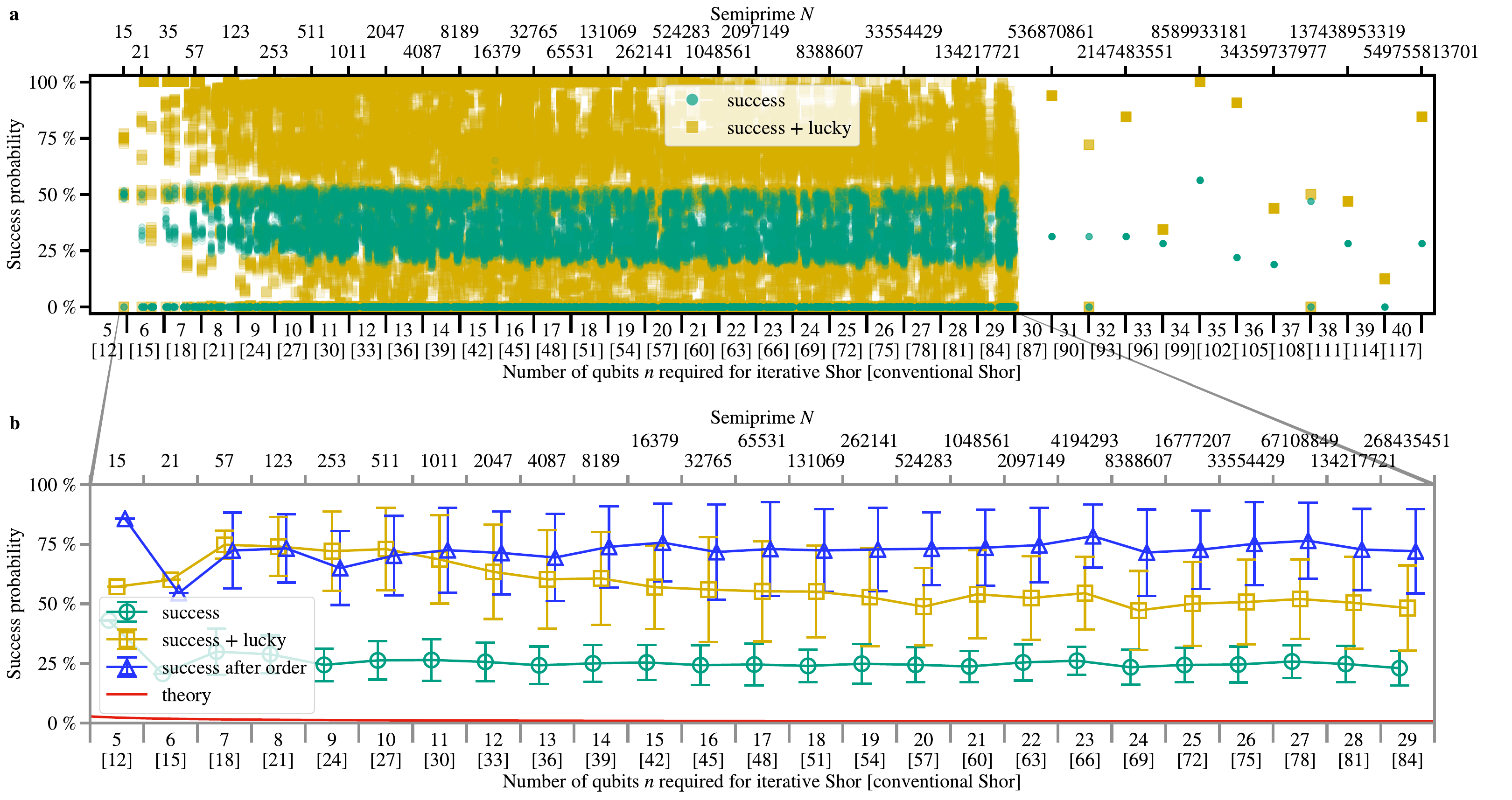}
  \caption{\textbf{Success probabilities for Shor's factoring algorithm.} For each factoring problem $(N,a)$, the success probabilities are given by the ratio of all $M$ bitstrings (sampled from Shor's algorithm) that yield a factor. If a bitstring satisfies all conditions of Shor's (original) algorithm, it is counted as ``success'' (green circles). If the bitstring yields a factor, even if these conditions are \emph{not} met, it is counted as ``success$+$lucky'' (yellow squares). 
  \textbf{a} 
  Individual success probabilities for each of the 61362 factoring problems $(N,a)$. The markers are placed at the positions of the factored semiprime according to the top axis. The number of qubits required using both the iterative and the conventional Shor algorithm is indicated on the bottom axis. 
  \textbf{b}
  Average success probabilities for the 52077 uniform factoring problems, averaged over all problems $(N,a)$ for a given number of qubits $n$ (see text). Error bars indicate the root-mean-square deviations of the averages of $a$ across different semiprimes $N$ for the same $n$. Blue triangles represent the success probabilities for all factoring problems $(N,a)$ that can be solved after a bitstring has yielded the order of $a$ modulo $N$ (which always happened within the first 33 bitstrings, see Fig~\ref{fig:histograms}b). The red line represents the theoretical bound for Shor's post-processing, given by 
  $2e^{-\gamma}/\pi^2\log\log N$
  (see Appendix~\ref{app:probabilitytheory}).
  Lines are guides to the eye.}
  \label{fig:results}
\end{figure*}

\subsection{Details on Memory and Computing Time}
\label{sec:memorycomputingtime}

The largest part of the memory needed by \texttt{shorgpu} is taken by the coefficients of the statevector $\ket\psi$ (see Eq.~(\ref{eq:psicoefficients})). For a 40-qubit iterative Shor circuit, which can be used to factor 39-bit integers, the statevector needs $2^{40}$ complex double-precision floating point numbers, so $16\times2^{40}\,{\mathrm{B}} = 16\,{\mathrm{TiB}}$. For performance reasons, two statevector buffers are used in the implementation of the oracle gate and the following single-qubit gates. 
In addition to the two statevector buffers, \texttt{shorgpu} requires two 32-bit integer buffers for the implementation of the oracle gate, called \texttt{oracle\_idxrecv} and \texttt{oracle\_idxsend} (see above). Each of these takes another $4\times2^{40}\,{\mathrm{B}} = 4\,{\mathrm{TiB}}$. The total GPU memory required is thus slightly larger than $40\,\mathrm{TiB}$. When using $N_{\mathrm{GPU}}=2048$ GPUs, the required memory per GPU is slightly larger than $20\,\mathrm{GiB}$.

We performed all simulations on JUWELS Booster~\cite{JUWELS,JuwelsClusterBooster}, a GPU cluster with 3744 NVIDIA A100 Tensor Core GPUs~\cite{nvidiaa100}, each of which has $40\,\mathrm{GiB}$ of GPU memory. Note that the implementation of the algorithm requires the number of GPUs to be a power of two (cf.~Section~\ref{sec:initialization}), so the maximum number of NVIDIA A100 GPUs that we can use on JUWELS Booster is 2048. The total computing time used to  perform the simulations amounts to 594 core years (corresponding to 49.5 GPU years since each node contains 4 A100 GPUs and 48 physical CPU cores).
We note that the total computing time is $22\,\%$ of the 2700 core years used for the recent factoring record of RSA-250---a number with 829 binary digits from the famous RSA factoring challenge~\cite{Boudot2020FactoringRSA250}.

\section{Results}
\label{sec:3}

In this section, we describe and interpret the results obtained from simulating Shor's algorithm according to Fig.~\ref{fig:scheme}. For our analysis, we generated 61362 factoring problems $(N,a)$, 52077 of which were chosen to have uniformly distributed prime factors to ensure unbiased results, and the rest comprise individual factoring problems for large semiprimes (see Appendix~\ref{app:problems}). We consider Shor's original post-processing in Section~\ref{sec:postprocessingshor} and Eker\aa's post-processing in Section~\ref{sec:postprocessingekera}.

\subsection{Using Shor's Post-Processing}
\label{sec:postprocessingshor}

\begin{figure*}
  \centering
  \includegraphics[width=\textwidth]{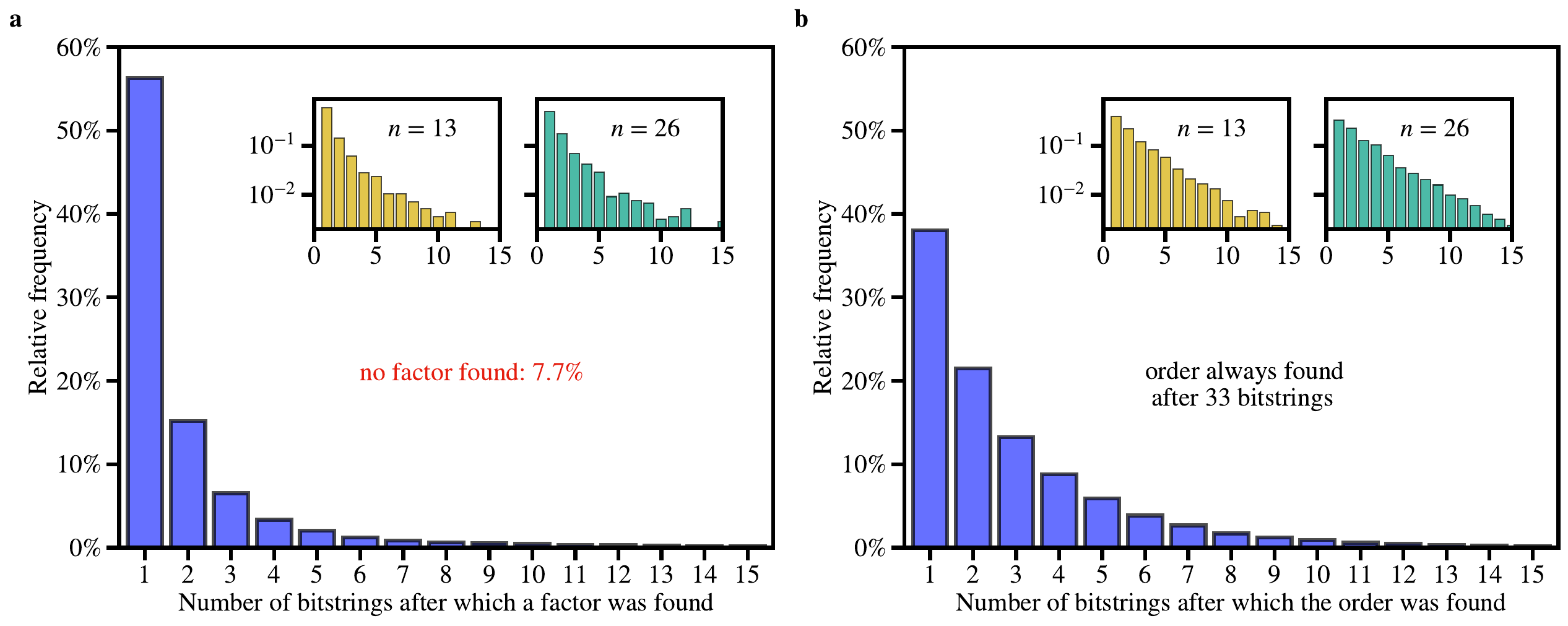}
  \caption{\textbf{Statistical analysis of how often Shor's algorithm has to be executed to solve a factoring problem.} Shown is the number of bitstrings that had to be generated until \textbf{a}, a factor of $N$, or \textbf{b}, the order of $a$ modulo $N$ could be found. The main plots show the statistics extracted from all 52077 uniformly distributed factoring problems $(N,a)$, for which the total number of sampled bitstrings is $M=1024$. Insets show the same information (on a logarithmic scale) for subsets of 2500 problems that all require the same number of qubits $n=13$ and $n=26$ for the iterative Shor algorithm (corresponding to $n=36$ and $n=75$ for the conventional Shor algorithm, respectively).}
  \label{fig:histograms}
\end{figure*}

\begin{figure*}
  \centering
  \includegraphics[width=\textwidth]{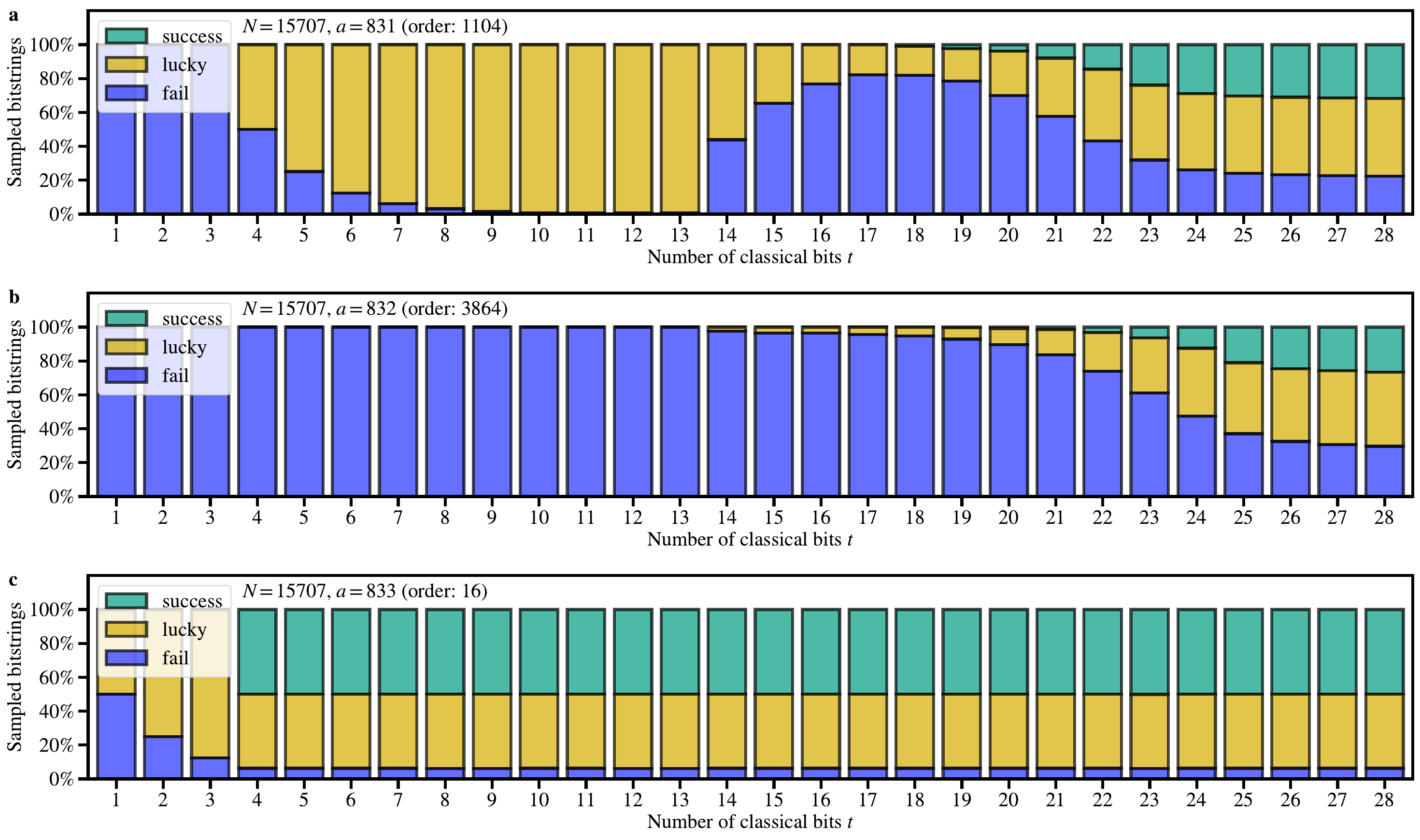}
  \caption{\textbf{Success probabilities of Shor's algorithm when less than the recommended $t=\lceil2\log_2N\rceil$ classical bits are extracted from the QFT.} Every bar represents the fraction of 1 million sampled bitstrings classified as ``success'' (green), ``lucky'' (yellow), and ``fail'' (blue). 
  The factored semiprime is $N=15707$ (such that $t=28$) with \textbf{a} $a=831$, \textbf{b} $a=832$, and \textbf{c} $a=833$. The corresponding orders of $a$ modulo $N$ are indicated at the top. Increasing $t$ beyond $28$ does not further improve the success probabilities.}
  \label{fig:scalet}
\end{figure*}

A given factoring problem for Shor's algorithm consists of a semiprime $N$ and a random integer $1<a<N$ coprime to $N$.
For each such factoring problem $(N,a)$, Shor's algorithm produces a sample of $M$ bitstrings (we typically consider $M=1024$ samples). Each bitstring $j$ is analyzed using the so-called standard procedure (see Appendix~\ref{app:standardprocedure}). If all checks on $j$ from the standard procedure pass, the algorithm was successful and we count the bitstring $j$ as ``success''. 
However, if certain checks on $j$ fail, we \emph{still} evaluate $j$ and test if it yields a factor of $N$. If it does, we count this factoring attempt as ``lucky''. Figure~\ref{fig:results}a shows a scatter plot of all ``success'' and ``success+lucky'' probabilities for all uniformly distributed problems ($n<30$ qubits) and the individual large problems ($30\le n \le40$ qubits).

Surprisingly, ``lucky'' occurs much more often than expected. In Fig.~\ref{fig:results}b, we see that on average only $25\,\%$ of all bitstrings yield ``success''. Including the ``lucky'' cases, however, a factor of the semiprime $N$ can be extracted from over $50\,\%$ of all bitstrings on average. Additionally, all average success probabilities are significantly larger than the theoretical bound of $3$--$4\,\%$ (see Appendix~\ref{app:probabilitytheory}). We conjecture that asymptotically, the average success probability for ``success$+$lucky'' approaches $50\,\%$ from above (further evidence is given in Section~\ref{sec:luckyclassification} below, where we give a classification of the different ``lucky'' scenarios and show that the main contribution saturates around $25\,\%$). This observation is remarkable, as it shows that factoring a semiprime with Shor's algorithm is often successful, even though the order-finding procedure actually fails. 

Simulating Shor's algorithm for semiprimes $N$ between 536870861 and 549755813701 requires substantial computational resources. Therefore, only individual cases are shown in Fig.~\ref{fig:results}a. These cases correspond to the largest ``interesting'' semiprimes for a given number of qubits $n=30,\ldots,40$. A noteworthy case is the factoring problem for $(N,a)=(8589933181,3974323683)$ ($n=34$ qubits). Here, the ``lucky'' cases raise the success probability from $56.25\,\%$ to $100\,\%$ (yellow square) among all $M$ bitstrings. Furthermore, the factoring problem for $(N,a)=(274877906893,226009433972)$ ($n=39$ qubits, second from the right) has a success probability of $0\,\%$ when the sufficient conditions for Shor's algorithm are presupposed (green circle). However, when ignoring the violations of these conditions, we find that Shor's algorithm can indeed factor $N$ with a ``lucky'' success probability of $12.5\,\%$ (yellow squares).

The unexpectedly large success probabilities when the lucky cases are included prompt the question ``how many bitstrings do we need to sample until a factor is found?'' This is a relevant question, since for large problems, computing time on both classical and quantum computers is an essential resource.
Figure~\ref{fig:histograms}a demonstrates that, for more than half of all factoring problems examined, the first sampled bitstring already yields a factor of $N$. Furthermore, in only $7.7\,\%$ of all factoring problems $(N,a)$, none of the $1024$ bitstrings produced a factor.
In this case, the reason is usually that the choice of $a$ was bad, which can be estimated to happen with probability 50\,\% (see proposition \textbf{C} in Appendix~\ref{app:probabilitytheory}). Clearly, the failure probability of $7.7\,\%$ is much smaller than the theoretical estimate would suggest.

Figure~\ref{fig:histograms}b further reveals that, even when the order-finding procedure in Shor's algorithm fails, the first bitstring often still produces a factor. In $38\,\%$ of all cases, the first bitstring yields the order of $a$ modulo $N$ (leftmost bar). From Fig.~\ref{fig:results}b, we know that on average, $75\,\%$ of all factoring problems can be solved after the order is known (blue triangles). Thus we expect approximately $38\,\%\times75\,\%\approx29\,\%$ of all factoring problems to be solved by the correct order after the first bitstring. However, in Fig.~\ref{fig:histograms}a, we see that $56\,\%$ of all problems are solved by processing the first bitstring. This percentage obviously is much larger than $29\,\%$, implying that it is easier to find a factor with Shor's algorithm than to solve the underlying order-finding problem.

Another interesting result is observed when reducing the number of bits $t$ in the sampled bitstring below the recommended $\lceil2\log_2N\rceil$ (cf.~Appendix~\ref{app:standardprocedure}). This saves resources in both versions of Shor's algorithm. For the conventional Shor algorithm, it reduces the required number of qubits and gates required for the QFT. For the iterative Shor algorithm, it linearly reduces the number of quantum gates and thus the execution time.

Surprisingly, in almost all cases, reducing the number of bits $t$ still allows for a successful factorization. Three representative cases are shown in Fig.~\ref{fig:scalet}. First, Fig.~\ref{fig:scalet}a shows that reducing $t$ may even increase the frequency of ``lucky'' factorizations to over $99\,\%$, as it does for $10\le t\le 13$ in this case. Second, in Fig.~\ref{fig:scalet}b, we see that even though the success probabilities decrease with $t$, at half of the recommended number of classical bits, that is at $t=14$, there are still ``lucky" cases, allowing for successful factorization. Finally, in the case shown in Fig.~\ref{fig:scalet}c, the success and lucky probabilities are essentially constant for $4\le t\le 28$. 

Although it is known that reducing $t$ may still allow for non-zero success probabilities~\cite{Barenco1996ApproximateQFT,Coppersmith2002ApproximateQFTForShor,Fowler2004ShorWithImpreciseRotationGates,Nam2012ShorPerformanceBandedQFT}, the surprising robustness (or even increase) of the ``lucky'' success probabilities has not been appreciated.
In conclusion, Shor's algorithm can still be successful (sometimes even more successful) if much less classical bits $t$ are sampled than the recommended $t=\lceil2\log_2N\rceil$.

\subsubsection{Classification of the ``Lucky'' Scenarios}
\label{sec:luckyclassification}

\begin{figure*}[b]
  \centering
  \includegraphics[width=\textwidth]{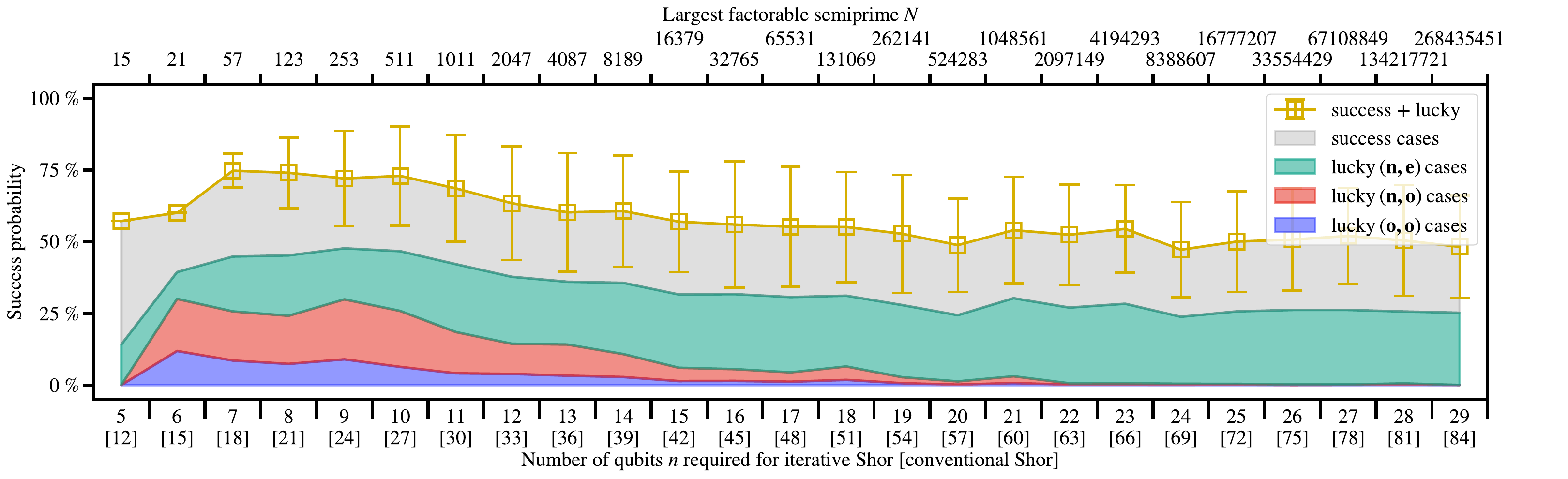}
  \caption{\textbf{Breakdown of the average success probabilities for successful factoring scenarios.} The contributions to the average ``success$+$lucky'' probability (yellow squares) from Fig.~\ref{fig:results}b are divided into the ``success'' cases (gray), the lucky (\textbf{n,e}) cases (green), the lucky (\textbf{n,o}) cases (red), and the lucky (\textbf{o,o}) cases (blue). Note that an $L$-bit semiprime requires $n=L+1$ qubits for the iterative Shor algorithm and roughly $3L$ qubits using the conventional Shor algorithm, as indicated on the bottom axis. The largest odd semiprime that can be factored with a given number of qubits is shown on the top axis. Lines are guides to the eye.
  }
  \label{fig:luckyresults}
\end{figure*}

\begin{figure*}[b]
  \centering
  \includegraphics[width=\textwidth]{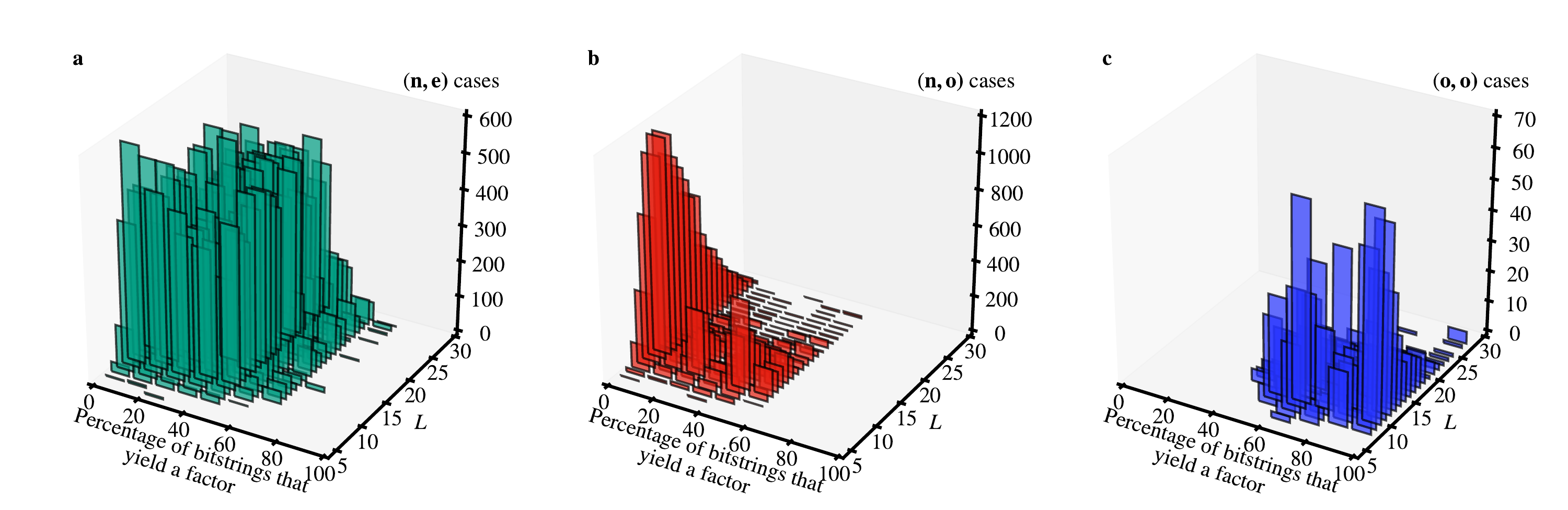}
  \caption{\textbf{Classification of the different ``lucky'' scenarios.} Shown is the absolute number of \textbf{a} (\textbf{n,e}) cases, \textbf{b} (\textbf{n,o}) cases, and \textbf{c} (\textbf{o,o}) cases, which yield a factor even though the sufficient conditions for Shor's algorithm are not met. 
  For each bit length $L=9,\ldots,29$, the total number of cases is given by 2500 uniformly distributed factoring problems (for smaller $L$, the total number is smaller than 2500 because all possibilities for factoring problems $(N,a)$ are exhausted, see~Appendix~\ref{app:problems}).
  Every bar represents a $10\,\%$-wide half-open interval $(P-10\,\%,P]$ with $P=10\,\%,20\,\%,\ldots,100\,\%$ for the percentage of all $M=1024$ bitstrings that yield a factor in this case. For instance, the large leftmost red bar at $L=11$ and $P=10\,\%$ in panel \textbf{b} means that in 1197 out of 2500 cases, up to $10\,\%$ of all sampled bitstrings yield a factor even though they represent an (\textbf{n,o}) case, i.e., they produce an odd number $r$ that is not the order. Similarly, the large rightmost blue bar at $L=9$ in panel \textbf{c} means that in 68 out of 2500 cases, more than $90\,\%$ of all sampled bitstrings yield a factor even though the correctly extracted order $r=\hat r$ is odd.
  }
  \label{fig:luckybars}
\end{figure*}

If a sampled bitstring $j$ does not pass the standard tests required by Shor's algorithm, but still produces a factor with the procedure shown in Fig.~\ref{fig:scheme}, we call this a ``lucky'' case. As shown above, this happens much more often than expected. In this section, we explain and classify the different scenarios that can happen. 

For a given factoring problem with an $L$-bit semiprime $N$ and an integer $a$ coprime to $N$, let $\hat r$ be the multiplicative order of $a$ modulo $N$. Furthermore, let $r$ be the denominator and $k$ be the numerator extracted from the convergent $k/r$ to $j/2^t$ using the continued fractions algorithm from the standard procedure (i.e., the largest $r<N$ such that $k/r$ is a convergent to $j/2^t$ with $|k/r-j/2^t|\le1/2r^2$~\cite{shor1997algorithm}). 
We distinguish between three scenarios in which the standard checks for Shor's algorithm fail:
\begin{itemize}
    \setlength{\itemindent}{2em}
    \item[(\textbf{n,e})] $r\neq \hat r$ is \textbf{not} the order but $r$ is \textbf{even},
    \item[(\textbf{n,o})] $r\neq \hat r$ is \textbf{not} the order and $r$ is \textbf{odd},
    \item[(\textbf{o,o})] $r=\hat r$ is the \textbf{order} but the order is \textbf{odd}.
\end{itemize}
In Fig.~\ref{fig:luckyresults}, we show a breakdown of the average success probability for these scenarios. We see that the (\textbf{n,e}) scenario makes up a large fraction of the successful factorizations and its relevance grows for larger integers. In contrast, the (\textbf{n,o}) and (\textbf{o,o}) scenarios, where the extracted $r$ is odd, only matter for smaller integers. We explain the reasons for this in the discussions of each individual scenario below.

Figure~\ref{fig:luckybars} shows the number of cases for each scenario among the 52077 uniformly drawn factoring problems. We see that the cases where bitstrings yield a factor in the (\textbf{n,e}) scenario are responsible for a significant fraction of all successful factorizations. Indeed, as Fig.~\ref{fig:luckyresults} suggests, the relevance of this scenario also grows on average and tends to saturate above $25\,\%$ for larger $L$. We expect that this contribution persists for even larger semiprimes.

Contributions from the (\textbf{n,o}) and (\textbf{o,o}) scenarios seem to be responsible only for a small number of successful factorizations, and mostly only for small semiprimes up to $L=10$. It is remarkable, however, that for many factoring problems that can be factored in the (\textbf{o,o}) scenario, $50$--$100\%$ of all bitstrings yield a factor (see Fig.~\ref{fig:luckybars}c). 

To understand these effects, we discuss the different scenarios individually. The goal is to obtain an understanding for why the different scenarios occur. The number-theoretic ideas are similar to the algorithm used in~\cite{Ekera2021OnCompletelyFactoringAnyIntegerShor}, which goes back to Miller's algorithm~\cite[Lemma 5]{Miller1976ThesisPrimalityTestRandomizationFactoring}.

\subsubsection{The (\textbf{n,e}) Scenario}

From the quantum circuit of Shor's algorithm, one can compute the probability distribution for the bitstrings $j$ that are sampled at the measurement~\cite{Michielsen2005EventBasedSimulationOfUniversalQC} (see Appendix~\ref{app:distribution} for the derivation),
\begin{align}
    \label{eq:distribution}
    p_{\hat r,t}(j) &= \frac{\hat r}{2^{2t}} \left( 
        \frac
        {\sin(\pi \hat rj\lfloor \frac{2^t}{\hat r}\rfloor/2^t)}
        {\sin(\pi \hat rj/2^t)}
        \right)^2 \nonumber\\
    &+ \frac{2^t - \hat r\lfloor \frac{2^t}{\hat r}\rfloor}{2^{2t}}
        \frac
        {\sin( \pi \hat rj[2\lfloor \frac{2^t}{\hat r}\rfloor+1]/2^t)}
        {\sin(\pi \hat rj/2^t)},
\end{align}
where $\lfloor 2^t/\hat r\rfloor$ denotes the integral number of times that $\hat r$ fits into $2^t$.
Note that for a given factoring problem $(N,a)$ with $t$ classical bits per bitstring, this distribution only depends on the order $\hat r$ of $a$ modulo $N$. This is a consequence of the fact that the QFT in Shor's algorithm is used to determine the period of the function $f(k) = a^k\,\mathrm{mod}\,N$, which is exactly $\hat r$.

The distribution in Eq.~(\ref{eq:distribution}) is shown for a few representative cases in Fig.~\ref{fig:suppdistributions} in Appendix~\ref{app:distribution}. It is strongly peaked at $\hat r$ bitstrings (see also~\cite{Bourdon2007SharProbabilityEstimatesShor})
\begin{align}
    \label{eq:peakenumeration}
    j\in\{\mathrm{round}(\hat k\times2^t/\hat r)\,:\,\hat k=0,\ldots,\hat r-1\},
\end{align}
where $\hat k$ enumerates the $\hat r$ peaks. Given $j$, the continued fractions algorithm yields a convergent $k/r=\hat k/\hat r$ to $j/2^t$. 
However, if $\hat k$ and $\hat r$ have a common factor, the denominator $r$ from the extracted convergent will not be equal to the order $\hat r$. For instance, this is the reason that the ``success'' cases in Fig.~\ref{fig:results}a are typically below $50\,\%$, since every second peak corresponds to an even $\hat k$, and an even order $\hat r$ is a sufficient condition for success; hence, at least a factor of two is lost in $\hat k/\hat r$. We note that with a very small probability, this procedure may also yield an $r>\hat r$ if the sampled bitstring $j$ is not at one of the $\hat r$ peaks of $p_{\hat r,t}(j)$.

To understand why $r\neq \hat r$ may still yield a factor of $N$, we consider the case that the order $\hat r$ yields a factor of $N$ (as Fig.~\ref{fig:results}b shows, this case occurs with approximately $75\,\%$ frequency). In this case, we have 
\begin{align}
    \label{eq:findfactorp}
    \mathrm{gcd}(a^{\hat r/2}-1,N) &= p,\\
    \mathrm{gcd}(a^{\hat r/2}+1,N) &= q,
\end{align}
where $p$ and $q$ are the two prime factors of $N$. Let $2^d$ be the largest power of 2 in $\hat r$ such that $2\nmid \hat r/2^d$ (meaning $2$ does no divide $\hat r/2^d$). Note that often, $2^d\ge 4$ since the multiplicative order of the whole group $\mathbb Z_N^*$ is $\phi(N)=(p-1)(q-1)$, which is at least divisible by 4 (here, $\phi(N)=\vert Z_N^*\vert$ is Euler's totient function). In this case, Eq.~(\ref{eq:findfactorp}) can be written as
\begin{align}
    \label{eq:findfactorp4}
    \mathrm{gcd}((a^{\hat r/4}-1)(a^{\hat r/4}+1),N) &= p,
\end{align}
so either $(a^{\hat r/4}-1)$ or $(a^{\hat r/4}+1)$ contain the prime factor $p$. Since every second $\hat k$ in Eq.~(\ref{eq:peakenumeration}) is even, it is likely that $r\in\{\hat r/2,\hat r/4,\ldots,\hat r/2^{d-1}\}$ (each case decreasing in likelihood). If $r=\hat r/2$, Eq.~(\ref{eq:findfactorp4}) shows that by testing both $\mathrm{gcd}(a^{r/2}\pm1,N)$, a factor will be found. If $r=\hat r/4$, knowing that $r$ is even, we can further write Eq.~(\ref{eq:findfactorp4}) as 
\begin{align}
    \label{eq:findfactorp8}
    \mathrm{gcd}((a^{\hat r/8}-1)(a^{\hat r/8}+1)(a^{\hat r/4}+1),N) &= p,
\end{align}
so if $p$ does not happen to be in $(a^{\hat r/4}+1)$, also a factor will be found. This reasoning can be iterated up to the unlikely case that $r=\hat r/2^{d-1}$, where Eq.~(\ref{eq:findfactorp}) becomes
\begin{align}
    \label{eq:findfactorpall2}
    \mathrm{gcd}(&(a^{\hat r/2^d}-1)(a^{\hat r/2^d}+1)\nonumber\\
    &\times(a^{\hat r/2^{d-1}}+1)\cdots(a^{\hat r/4}+1),N) = p.
\end{align}
Similarly, if $r=\hat r/3$, we can write Eq.~(\ref{eq:findfactorp}) as
\begin{align}
    \label{eq:findfactorp3}
    \mathrm{gcd}((a^{r/2}-1)(a^r + a^{r/2} + 1),N) &= p,
\end{align}
in which case we also find a factor if $p$ happens to be in the first part of the product. Finally, if $r=\hat r/l$ with $l>4$, we can write Eq.~(\ref{eq:findfactorp}) as
\begin{align}
    \label{eq:findfactorpl}
    \mathrm{gcd}(&(a^{r/2}-1)\nonumber\\
    &\times(a^{r/2\times (l-1)} + a^{r/2\times(l-2)} + \cdots + a^{r/2} + 1),N) = p.
\end{align}
Note that as $l$ grows, this case becomes increasingly unlikely since $l$ would need to be a factor of $\hat k$ already. But also in this case, there is a small chance that when evaluating $\mathrm{gcd}(a^{r/2}-1,N)$, a factor can be found. We remark that when $r/2=\hat r/2l$ is prime, the decomposition in Eq.~(\ref{eq:findfactorpl}) is irreducible~\cite{HardyWright}, such that no further polynomial in $a^{r/2}$ including $p$ can be factored out.

\begin{figure*}
  \centering
  \includegraphics[width=\textwidth]{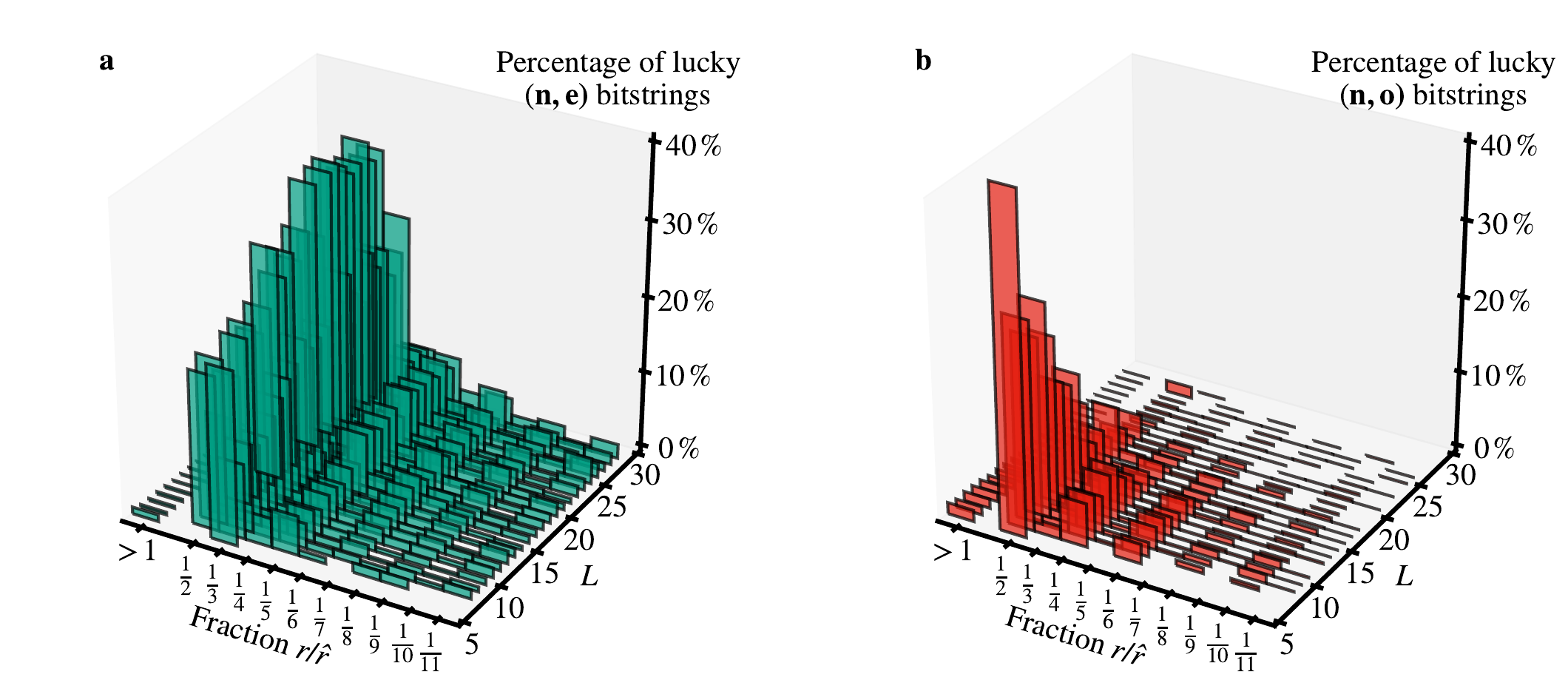}
  \caption{\textbf{Histograms of the fractions $r/\hat r$ between the order $\hat r$ and the denominator $r$ extracted from the continued fractions algorithm.} Shown is the percentage of bitstrings for $L\ge5$ (normalized by all ``lucky'' bitstrings) that yield a factor in the \textbf{a} (\textbf{n,e}) and \textbf{b} (\textbf{n,o}) scenario, respectively. Here, $r$ is not the order but often a large divisor of it. Additionally, the bars at the very left for $r/\hat r > 1$  represent the number of cases where the extracted $r$ is actually \emph{larger} than the order $\hat r$, which may occasionally happen when a bitstring $j$ is sampled at a non-zero probability $p_{\hat r,t}(j)$ that is not a peak (cf.~Figs.~\ref{fig:suppdistributions}b and c).
  }
  \label{fig:luckydivisors}
\end{figure*}

The distribution of the fractions $r/\hat r$ (extracted from the data generated from the uniformly distributed factoring problems) is shown in Fig.~\ref{fig:luckydivisors}a. Indeed, we see that very often a small multiple of $r$ is equal to the order $\hat r$. In particular, the significance of fractions up to $r/\hat r=1/11$ seems to increase with $L$, i.e., with increasingly large semiprimes $N$. This observation agrees well with the argument given in~\cite{Ekera2021OnCompletelyFactoringAnyIntegerShor}.

Interesting examples for the lucky (\textbf{n,e}) scenario are the individual problems for $n=34$ and $n=39$ discussed in Section~\ref{sec:postprocessingshor}. In particular, the $n=39$ case with $N=274877906893=364303\times754531$, $a=226009433972$ and order $\hat r=45812798010$ violates the condition $a^{\hat r/2}\not\equiv-1\ (\mathrm{mod}\,N)$. As this is one of the sufficient conditions for Shor's algorithm to guarantee successful factorization, the corresponding ``success'' probability is zero (green circle). However, $12.5\,\%$ of the sampled bitstrings yield even integers $r\in\{\hat r/3,\hat r/5,\hat r/111\}$, which still allow for a successful ``lucky'' factorization of $N$ (the corresponding quadratic residues $a^r$ do not have a trivial square root, i.e., $a^{r/2}\not\equiv-1\ (\mathrm{mod}\,N)$).

For large $N$, the (\textbf{n,e}) scenario makes up the majority of all ``lucky'' factorizations (see Fig.~\ref{fig:luckyresults}). We conjecture that on average, the probability of ``success$+$lucky'' factorizations asymptotically approaches $50\,\%$ due to the (\textbf{n,e}) scenario. 

\subsubsection{The (\textbf{n,o}) Scenario}

\begin{figure*}
  \centering
  \includegraphics[width=\textwidth]{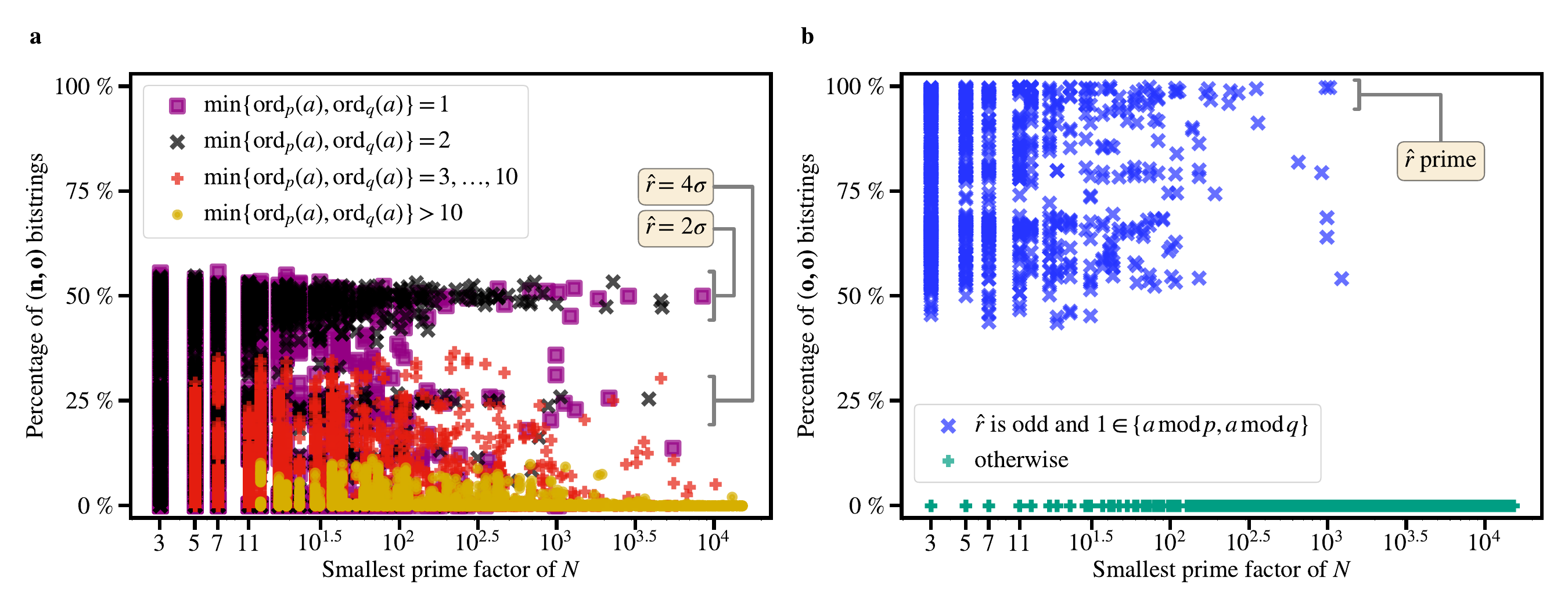}
  \caption{\textbf{Classification of ``lucky'' factorizations for the case that the extracted integer $r$ is odd, using $\lfloor r/2\rfloor=(r-1)/2$ instead of $r/2$.} Shown is the percentage of bitstrings that yield a factor as a function of the smallest prime factor of the semiprime $N=p\times q$.
  \textbf{a} The lucky (\textbf{n,o}) probabilities, classified in terms of the minimum order of $a$ modulo $p$ and $q$ (see legend; prioritized from top to bottom).
  \textbf{b} The lucky (\textbf{o,o}) probabilities, fully classified in terms of the cases where  $1\in\{a\,\mathrm{mod}\,p,a\,\mathrm{mod}\,q\}$ and $\hat r$ is odd (blue crosses) and the rest (green plusses). 
  In both \textbf{a} and \textbf{b}, annotations indicate an additional condition that is satisfied by many (but not all) cases at this percentage level ($\sigma$ denotes an odd integer). We use the notation $\mathrm{ord}_p(a)$ to denote the order of $a$ modulo $p$, i.e., the smallest integer $m$ such that $a^m\equiv1\quad(\mathrm{mod}\,p)$.
  }
  \label{fig:luckyrodd}
\end{figure*}

In the (\textbf{n,o}) scenario, the bitstring $j$ yields an integer $r$ that is neither the order $\hat r$ nor even.
Using the reasoning from the (\textbf{n,e}) scenario, this happens only if $r\mid \hat r/2^d$, so all $d$ powers of 2 must have been in $\hat k$ from Eq.~(\ref{eq:peakenumeration}) already (or $d=0$, in which case $\hat r$ is odd). This is an unlikely scenario given that all $\hat k\in\{0,\ldots,\hat r-1\}$ occur with roughly the same probability, see Fig.~\ref{fig:suppdistributions} in Appendix~\ref{app:distribution}. Moreover, as Fig.~\ref{fig:luckybars} shows, the frequency of this scenario tends to zero for larger semiprimes $N$. However, it does occur for smaller semiprimes with up to $25\,\%$ frequency on average (see the red area in Fig.~\ref{fig:luckyresults}), so it is instructive to understand how a factor can be found in this case. In what follows, we exclude the irrelevant case $r=1$ as it will never yield a factor.

Let $r=\hat r/l$ where $2^d\mid l$ and $2\nmid r$ (note that $d=0$ is possible if the order $\hat r$ is odd, but we always have $l\ge2$ since $r\neq \hat r$). In this case, using the procedure depicted in Fig.~\ref{fig:scheme}, we test
\begin{align}
   \label{eq:roddcomputefactor}
   \mathrm{gcd}(a^{\lfloor r/2\rfloor}\pm1,N) = \mathrm{gcd}(a^{(\hat r/l-1)/2}\pm1,N) 
\end{align}
to find a factor of $N$.

Since Eq.~(\ref{eq:roddcomputefactor}) seems somewhat arbitrary from the perspective of the original theory behind Shor's algorithm, one might think that a factor may only be found by coincidence. For instance, when $N=p\times q$ has a very small prime factor (say 3, 5, or 7), then whatever number is computed by Eq.~(\ref{eq:roddcomputefactor}) might have a chance of including the small prime factor. 

That this reasoning does not always hold is shown in Fig.~\ref{fig:luckyrodd}a, where we list the percentage of bitstrings that yield a factor in the (\textbf{n,o}) scenario as a function of the smallest prime factor of $N$. Indeed, we see that also larger prime factors can be found in certain cases. The most important of these are the cases in which $a$ has a small order with respect to either $p$ or $q$ (purple squares, black crosses, and red plusses in Fig.~\ref{fig:luckyrodd}a). Indeed, one can prove that if
\begin{align}
    \label{eq:conditionno}
    \mathrm{min}\{\mathrm{ord}_p(a),\mathrm{ord}_q(a)\}\in\{1,2\},
\end{align}
then $\mathrm{gcd}( a^{\lfloor r/2\rfloor} \pm 1,N )$ with $r\neq \hat r$ odd yields a factor of $N$.
\\~\\
\textbf{Proof:} 
Without loss of generality, we assume $\mathrm{ord}_p(a)\in\{1,2\}$.
This implies that $a\equiv \pm1\, (\mathrm{mod}\, p)$ (because if $\mathrm{ord}_p(a)=1$, we have $a\equiv 1\, (\mathrm{mod}\,p)$, and if $\mathrm{ord}_p(a)=2$, we have $a\equiv -1\, (\mathrm{mod}\,p)$). 
Hence,
\begin{align}
    x_p := a^{\lfloor r/2\rfloor} \,\mathrm{mod}\,p = (\pm1)^{\lfloor r/2\rfloor}\,\mathrm{mod}\,p = \pm1.\label{eq:sub_proof_case_mod_p}
\end{align}
Moreover, since $\hat r=\mathrm{lcm}( \mathrm{ord}_p(a),\mathrm{ord}_q(a))$, where $\mathrm{lcm}$ denotes the least common multiple, $\mathrm{ord}_p(a)\in\{1,2\}$ implies that $\mathrm{ord}_q(a) \in\{\hat r,\hat r/2\}$.

Thus we have $\mathrm{ord}_q(a)\ge \hat r/2 \ge \hat r/l > \hat r/l-1$ (where $l\ge2$ is defined above Eq.~(\ref{eq:roddcomputefactor})), and therefore
\begin{align}
    x_q := a^{\lfloor r/2\rfloor} \,\mathrm{mod}\,q = a^{(\hat r/l-1)/2}\,\mathrm{mod}\,q\neq \pm 1\label{eq:sub_proof_case_mod_q}
\end{align} 
(since otherwise $\hat r/l-1$ would be a multiple of the order of $a$ modulo $q$).
Applying the Chinese remainder theorem~\cite{HardyWright} to Eqs.~(\ref{eq:sub_proof_case_mod_p}) and~(\ref{eq:sub_proof_case_mod_q}), we obtain
\begin{align}
     a^{\lfloor r/2\rfloor} \equiv x_q p p_q + x_pqq_p\quad(\mathrm{mod}\,N),
\end{align}
where $p_q$ is the inverse of $p\ (\mathrm{mod}\,q)$ and $q_p$ is the inverse of $q\ (\mathrm{mod}\,p)$. Using that $pp_q+qq_p \equiv 1\ (\mathrm{mod}\,N)$, we finally have
\begin{align}
    \label{eq:sub_proof_no}
    \mathrm{gcd}( a^{\lfloor r/2\rfloor} \pm 1,N )
    &= \mathrm{gcd}( (x_q \pm 1) p p_q + (x_p \pm 1) qq_p, N ).
\end{align}
Thus, since $x_q\neq\pm1$, the case where $x_p\pm1=0$ will yield the factor $p$. $\hfill\square$
\\~\\
One can estimate how often the situation given by Eq.~(\ref{eq:conditionno}) happens. Choosing $a$ uniformly at random from $\{2,\ldots,N-1\}$ with $\mathrm{gcd}(a,N)=1$ is equivalent to choosing $a\,\mathrm{mod}\,p$ ($a\,\mathrm{mod}\,q$) uniformly at random from $\{1,\ldots,p-1\}$ ($\{1,\ldots,q-1\}$) with the exception of $a\,\mathrm{mod}\,p=a\,\mathrm{mod}\,q=1$. Thus there are $(p-2)(q-2)-1\sim pq$ choices for $a$, $2p+2q-13\sim p+q$ of these satisfying either $a\,\mathrm{mod}\,p=\pm1$ or $a\,\mathrm{mod}\,q=\pm1$. Therefore, the contribution of cases with $\mathrm{min}\{\mathrm{ord}_p(a),\mathrm{ord}_q(a)\}=1,2$ becomes negligible for large $p$ and $q$. We remark that individual cases with $\mathrm{min}\{\mathrm{ord}_p(a),\mathrm{ord}_q(a)\}=3,4,\ldots,$ may further contribute to lucky (\textbf{n,o}) factorizations, as Fig.~\ref{fig:luckyrodd}a shows.

\subsubsection{The (\textbf{o,o}) Scenario}

In the (\textbf{o,o}) scenario, the bitstring $j$ yields the order $r= \hat r$ and the order is odd. Interestingly, as Fig.~\ref{fig:luckyrodd}b suggests, this case can be fully classified, viz.~we can prove that $\mathrm{gcd}( a^{\lfloor r/2\rfloor} - 1,N )$ yields a factor if and only if $1\in\{a\,\mathrm{mod}\,p,a\,\mathrm{mod}\,q\}$ (see also~\cite{MartinLopez2012ShorExperimentQubitRecycling,Grosshans2015FactoringSafeSemiprimesShor,Lawson2015OddOrdersShor,Johnston2017OddOrdersShor}). Note that $a>1$ by construction, so one of the two is larger than 1, and in particular $r>1$.
\\~\\
\textbf{Proof:} 
The ``$\Leftarrow$'' case is a special case of the proof in the (\textbf{n,o}) scenario. If $a\,\mathrm{mod}\,p=1$, we have $x_p=1$ in Eq.~(\ref{eq:sub_proof_case_mod_p}) and $x_q\neq\pm1$ in Eq.~(\ref{eq:sub_proof_case_mod_q}), so $\mathrm{gcd}( a^{\lfloor r/2\rfloor} - 1,N )=p$ by Eq.~(\ref{eq:sub_proof_no}).
We therefore only need to show the ``$\Rightarrow$'' case. Without loss of generality, let $\mathrm{gcd}( a^{\lfloor r/2\rfloor} - 1,N )=p$, so $p\mid (a^{\lfloor r/2\rfloor} - 1)$.
From this follows that $a^{\lfloor r/2\rfloor} \equiv 1\ (\mathrm{mod}\,p)$, so $\mathrm{ord}_p(a) \mid \lfloor r/2\rfloor = (r-1)/2$. However, we also have $\mathrm{ord}_p(a)\mid r$ because $r=\mathrm{lcm}(\mathrm{ord}_p(a),\mathrm{ord}_q(a))$. Since $\mathrm{gcd}((r-1)/2, r) = 1$ (using the Euclidean algorithm), this is only possible if $\mathrm{ord}_p(a) = 1$, which means $a\,\mathrm{mod}\,p=1$.  $\hfill\square$
\\~\\
Next we show that the ``$+$'' case, i.e.~$\mathrm{gcd}( a^{\lfloor r/2\rfloor} + 1,N )$, never gives a factor in the case $r=\mathrm{ord}_N(a)$ odd.
\\~\\
\textbf{Proof:} Assume $\mathrm{gcd}(a^{(r-1)/2}+1,N)=p$.
This means that $a^{(r-1)/2} \equiv -1 \ (\mathrm{mod}\, p)$, and thus $a^{r-1} \equiv 1 \ (\mathrm{mod}\, p)$.
The former implies that $\mathrm{ord}_p(a) \, \nmid\, (r-1)/2$ and the latter implies that $\mathrm{ord}_p(a) \mid r-1$.
Therefore, $2 \mid \mathrm{ord}_p(a)$. From $r=\mathrm{lcm}(\mathrm{ord}_p(a),\mathrm{ord}_q(a))$ then follows that $2 \mid r$, but this is a contradiction because $r$ was assumed to be odd. $\hfill\square$
\\~\\
Finally, we can show that it does not matter whether we round $r/2$ up or down, i.e., whether we take $\lfloor r/2\rfloor=(r-1)/2$ or $\lceil r/2\rceil=(r+1)/2$ in Fig.~\ref{fig:scheme}, since one of them yields a factor whenever the other one also yields a factor:
\begin{align}
   \mathrm{gcd}(a^{\lfloor r/2\rfloor}-1,N) &= \mathrm{gcd}(a^{\lfloor r/2\rfloor}(a^{\lceil r/2\rceil} - 1),N) \nonumber\\
   &= \mathrm{gcd}(a^{\lceil r/2\rceil} - 1,N),
\end{align}
where we used that $a^{\lfloor r/2\rfloor}a^{\lceil r/2\rceil}\equiv a^{(r-1)/2}a^{(r+1)/2}\equiv a^r\equiv 1\ (\mathrm{mod}\,N)$ and furthermore that $a^{\lfloor r/2\rfloor}$ cannot have a common factor with $N$ (since $a$ was chosen coprime to $N$),

A special, additional condition that yields a success probability of almost $100\,\%$ with the (\textbf{o,o}) scenario is when $\hat r$ is prime, as indicated in Fig.~\ref{fig:luckyrodd}b. In this case, almost all bitstrings $j$ are sampled at the peaks of Shor's bitstring distribution $p_{\hat r,t}(j)$ given by Eq.~(\ref{eq:peakenumeration}) and directly yield the order $r=\hat r$, because $\hat k$ and $\hat r$ are always coprime. The only exceptions are either when $j$ belongs to the first peak corresponding to $\hat k=0$, or when $j$ lies in the neighborhood of one of the peaks of Eq.~(\ref{eq:distribution}) where the probability is small.

\subsection{Using Eker\aa's Post-Processing}
\label{sec:postprocessingekera}

In 2022, Eker{\aa} has proven a lower bound for the success probability that takes into account additional, efficient classical post-processing procedures~\cite{Ekera2022OnTheSuccessProbabilityOfQuantumOrderFindingShor} (his implementation of the procedures can be found in~\cite{Ekera2023Quppy}).
While with Shor's post-processing, a factor is found with more than $50\,\%$ probability on average after a single run (see Fig.~\ref{fig:histograms}a), with Eker{\aa}'s post-processing, it is possible to increase this probability arbitrarily close to unity.
The bound reads
\begin{widetext}
\begin{align}
    \label{eq:shorboundekera}
    p(\text{success}\mid N) \ge 
    \underbrace{
    \bigg(1-\frac{1}{\pi^2}\bigg(\frac 2 B + \frac 1 {B^2} + \frac 1 {3B^2}\bigg) - \frac{\pi^2(2B+1)}{\sqrt{2^{m+\ell}}}\bigg)
    }_{\text{$j$ sampled $\pm B$ bitstrings around peak}}
    \underbrace{
    \bigg(1 - \frac 1 {c\log cm}\bigg)
    }_{\text{peak yields order $\hat r$}}
    \underbrace{
    \bigg(1 - 2^{-k}{n_F\choose2} - \frac 1 {2\varsigma^2\log^2\varsigma L}\bigg)
    }_{\text{order $\hat r$ yields factors of $N$}}
    ,
\end{align}
\end{widetext}
where $L$ is the bit length of $N$, $m+\ell=t$ is the number of classical bits obtained from Shor's algorithm, 
$n_F$ is the number of distinct prime factors of $N$ (i.e., $n_F=2$ for semiprimes), and $B, c, k, \varsigma\ge 1$ are constants of the post-processing algorithms that can be freely selected.
We choose $m=L$ and $\ell=t-L$ so that the results are in line with the analysis presented above. Note that the only technical requirement is $2^m>\hat r$ and $2^{m+\ell}>\hat r^2$, so $m=L-1$ is possible~\cite{Ekera2022OnTheSuccessProbabilityOfQuantumOrderFindingShor}; this does not make a difference for the results that follow.

We remark that the three factors in Eq.~(\ref{eq:shorboundekera}) are directly related to the three propositions \textbf{A}, \textbf{B}, and \textbf{C} discussed in Appendix~\ref{app:probabilitytheory}. We discuss each factor in turn.

The first factor in Eq.~(\ref{eq:shorboundekera}) comes from the idea that whenever the bitstring $j$ is not sampled at one of the $\hat r$ peaks (see Fig.~\ref{fig:suppdistributions} in Appendix~\ref{app:distribution}), it is often sampled very close to a peak. Thus, one can try out all bitstrings in the range $\{j-B, \ldots, j+B\}$ for some small $B$. The probability to find the peak among these bitstrings can be estimated from the distribution $p_{\hat r,t}(j)$ in Eq.~(\ref{eq:distribution}). Instead of $4/\pi^2$ (see Eq.~(\ref{eq:propositiona}); this would correspond to $B=0$), we then get a larger probability, given by the first factor. Here we choose $B=L$, i.e.~the number of bits in $N$.

The second factor in Eq.~(\ref{eq:shorboundekera}) stems from the idea that, when the continued fraction method does not yield the order $\hat r$, it will often yield a large divisor $r = \hat r/D$ for some small $D$ (see Fig.~\ref{fig:luckydivisors}). Starting from $r$, Eker{\aa} gives several classical algorithms in~\cite{Ekera2022OnTheSuccessProbabilityOfQuantumOrderFindingShor} to efficiently recover the real order $\hat r = r\times D$. The corresponding success probability is given by $1 - 1/c\log cm$, where $c\ge1$ is a parameter that is free to choose. Its derivation is based on the probability that $D$ is $cm$-smooth, meaning that $D>0$ is not divisible by any prime power larger than $cm$. For our numerical work we choose $c=1$.

Finally, the third factor in Eq.~(\ref{eq:shorboundekera}) follows from the algorithm presented in~\cite{Ekera2021OnCompletelyFactoringAnyIntegerShor} (see also~\cite{Grosshans2015FactoringSafeSemiprimesShor}). This algorithm describes the factoring of an arbitrary composite integer $N$ (with $n_F\ge2$ distinct prime factors) given the order $\hat r$ of a single element $a\in\mathbb Z_N^*$ selected uniformly at random. The corresponding success probability depends on two parameters $k,\varsigma\ge1$ ($\varsigma$ is called $c$ in~\cite{Ekera2021OnCompletelyFactoringAnyIntegerShor}) that can be freely selected. For our numerical work we choose $k=100$ and $\varsigma=1$.

Probably the most important consequence of Eker{\aa}'s result given by Eq.~(\ref{eq:shorboundekera}) is that, as the size of the factoring problem becomes very large, i.e.~$N,\hat r\to\infty$, the success probability approaches one. This trend can already be seen in Fig.~\ref{fig:errors}a (gray line), which shows that the bound is increasing---even though it is already quite large for our modest choice of $(B,c,k,\varsigma)=(L,1,100,1)$. Furthermore, the gray diamonds in Fig.~\ref{fig:errors}a represent the actual success probabilities, obtained from applying Eker\aa's post-processing to the largest scenarios studied above. They are all larger than $93\,\%$ and thus even closer to unity than expected (this potential underestimation was noted in~\cite{Ekera2022OnTheSuccessProbabilityOfQuantumOrderFindingShor}).

\subsection{Errors During the Execution of Shor's Algorithm}

With Eker\aa's post-processing~\cite{Ekera2021OnCompletelyFactoringAnyIntegerShor,Ekera2022OnTheSuccessProbabilityOfQuantumOrderFindingShor}, the expected success probability using only a single run of the quantum part of the algorithm can be brought arbitrarily close to $100\,\%$ by properly selecting the constants $(B,c,k,\varsigma)$ (cf.~Eq.~(\ref{eq:shorboundekera})). However, these probabilistic estimates still require a successful execution of the quantum part of Shor's algorithm. Since fully error-corrected, fault-tolerant quantum computers will probably not become available for several years to come~\cite{GoogleQuAI2021,Krinner2022,Sivak2023QuantumErrorCorrectionBreakEven}, it is an interesting, relevant question to study how the performance of the post-processing algorithms is affected by errors during the execution of the quantum algorithm.

\begin{figure*}
  \centering
  \includegraphics[width=\textwidth]{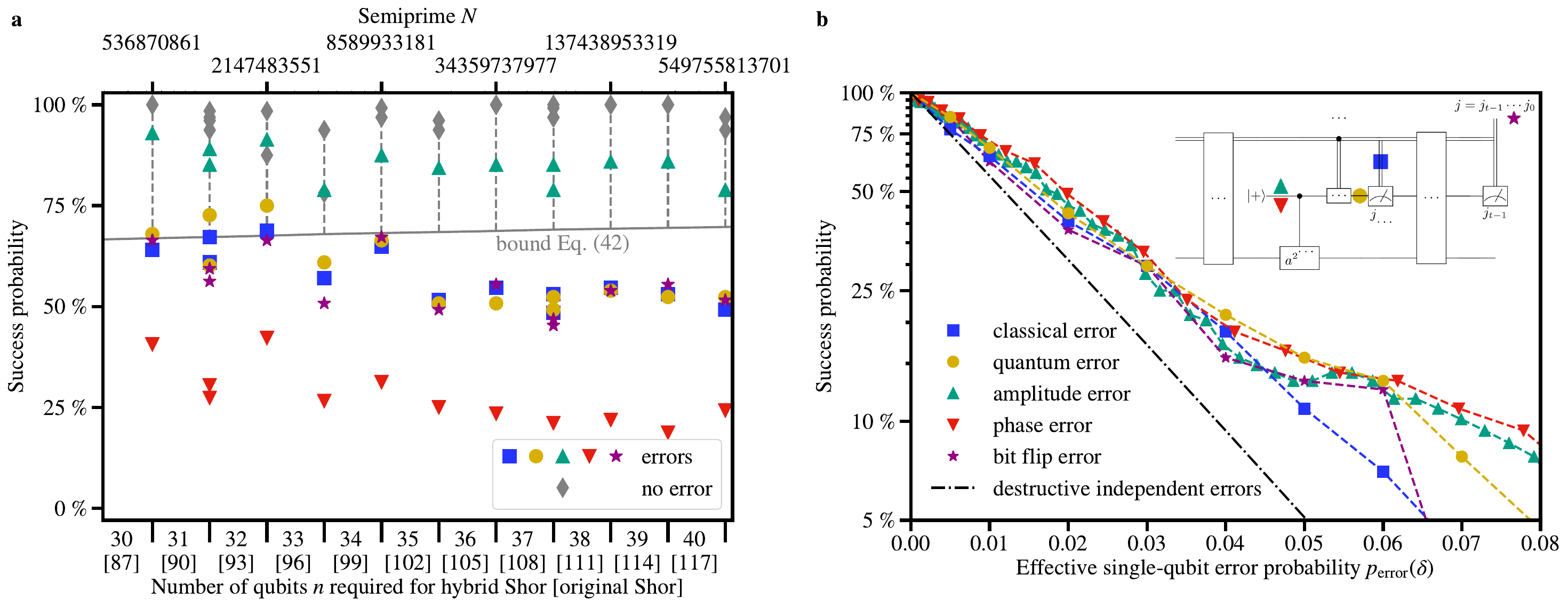}
  \caption{\textbf{Performance of Shor's algorithm using Eker\aa's post-processing in the presence of errors.} 
  \textbf{a} Success probability using Eker\aa's post-processing with $(B,c,k,\varsigma)=(L,1,100,1)$ for the largest scenarios from Fig.~\ref{fig:results}a in the case of classical measurement errors with $\delta=0.01$ (blue squares), quantum measurement errors with $\delta=0.01$ (yellow circles), amplitude initialization errors with $\delta=0.1$ (green upward-pointing triangles), phase initialization errors with $\delta=0.1$ (red downward-pointing triangles), bit flip errors (purple stars) with $\delta=0.01$, or no errors (gray diamonds). Additionally, the solid gray line shows the bound Eq.~(\ref{eq:shorboundekera}) for semiprimes with $n_F=2$ prime factors, and the vertical dashed gray lines show by how much the bound underestimates the actual performance (cf.~\cite{Ekera2022OnTheSuccessProbabilityOfQuantumOrderFindingShor}).
  \textbf{b} Success probability as a function of the effective single-qubit error probability $p_{\mathrm{error}}(\delta)$ for the 30-qubit case $N=536870903$ from panel a. The black dash-dotted line represents the success probability in the case of independent destructive errors, $(1-p_{\mathrm{error}})^t$. The inset shows schematically in which parts of the quantum circuit in Fig.~\ref{fig:shor} each of the different errors happen.
  }
  \label{fig:errors}
\end{figure*}

In this section, we consider five different models for errors arising in the quantum part of Shor's algorithm. Each of these is shown in the inset of Fig.~\ref{fig:errors}b, which schematically marks the places in the iterative Shor algorithm (cf.~Fig.~\ref{fig:shor}) at which the respective errors may occur.
\begin{enumerate}
    \item Classical measurement errors (blue squares) are defined as misclassifications occurring directly after each quantum measurement process with a constant error probability $p_{\mathrm{error}}(\delta)=\delta$ (see Section~\ref{sec:classicalerror}).
    \item Quantum measurement errors (yellow circles) are modeled as depolarizing quantum noise during the measurement process with effective error probability $p_{\mathrm{error}}(\delta)=\delta=p_x+p_y$ (see Section~\ref{sec:quantumerror}).
    \item Amplitude initialization errors (green upward-pointing triangles) are modeled by initializing the recycled qubit not in the uniform superposition $\ket+=(\ket0+\ket1)/\sqrt{2}$, but by increasing the amplitude of $\ket0$ as a function of $\delta$. The effective error probability is $p_{\mathrm{error}}(\delta)=(1-\sqrt{1-\delta^2})/2$ (see Section~\ref{sec:amplitudeerror}).
    \item Phase initialization errors (red down-pointing triangles) are defined by introducing a relative phase $e^{i\pi\delta}$ between the states $\ket0$ and $\ket1$ in the initialization. The effective error probability is $p_{\mathrm{error}}(\delta)=(1-\cos(\pi\delta))/2$ (see Section~\ref{sec:phaseerror}).
    \item Bit flip errors (purple stars) are defined by flipping each bit in the final bitstring $j$ with probability $p_{\mathrm{error}}(\delta)=\delta$. This error model, in contrast to the others, does not affect the execution of the quantum part of the iterative Shor algorithm. While such an error (e.g., a fault in the classical computer memory) may be considered unlikely, it is still interesting to compare its consequences to the errors in the quantum part.
\end{enumerate}
We consider the case that for each of these errors, Eker\aa's post-processing algorithm is applied to the resulting bitstrings, without the user being aware that one or more errors may have occurred.

Figure~\ref{fig:errors}a shows the success probabilities for the different errors and problem sizes. We see that errors with $\delta=0.01$ (which corresponds to $1\,\%$ error probability for the blue squares, yellow circles, and purple stars) can decrease the success probability below the bound Eq.~(\ref{eq:shorboundekera}) indicated by the solid gray line. Furthermore, the success probabilities show a decrease as a function of problem size that rivals the increasing success probability from the bound Eq.~(\ref{eq:shorboundekera}). Nevertheless, Eker\aa's post-processing algorithm still produces correct factors even in the presence of errors.

Figure~\ref{fig:errors}b shows the scaling of the success probability as a function of the effective single-qubit error probability for the 30-qubit case $N=536870903$. For all errors, we see that the performance of the factoring algorithm including Eker{\aa}'s post-processing scales similarly, despite the fundamental differences in the error models. This type of universal behavior is an interesting and unexpected observation.

It is also instructive to compare the simulation results to $(1-p_{\mathrm{error}})^t$ (black dash-dotted line in Fig.~\ref{fig:errors}b), which represents the probability to obtain a bitstring for which no error occurred (under the assumption that errors for individual bits are independent). Since we know from Fig.~\ref{fig:errors}a that the considered $30$-qubit case is solved with $100\,\%$ success, $(1-p_{\mathrm{error}})^t$ represents the assumption that errors are destructive, i.e., an error in one of the bits prevents a successful factorization. Hence, the systematic gap between the dash-dotted line and the other dashed lines in Fig.~\ref{fig:errors}b shows that the quantum factoring problem can still be solved with Eker{\aa}'s post-processing in the presence of errors. The fact that the success probability is systematically larger than the one for independent errors is encouraging, because an error at one stage in the iterative Shor algorithm affects the operation of all subsequent gates that depend on previous measurement results (see inset of Fig.~\ref{fig:errors}b). Such an error can thus propagate through the quantum algorithm and induce further, correlated errors. Our simulation results reveal a certain resilience when using Eker\aa's post-processing in combination with the iterative Shor algorithm for factoring integers.

\subsection{Discussion of Limitations and Future Directions}

Some of our design choices, made to achieve a large-scale simulation of Shor's algorithm for many factoring scenarios, result in certain practical limitations to what we can simulate. In this section, we list these design choices, state the accompanying limitations, and discuss interesting future research directions that alternative choices could offer.

\begin{enumerate}
    \item In a practical realization of Shor's quantum circuit shown in Figure~\ref{fig:shor}, most of the work is expected to be in the implementation of the exponentiation in terms of the controlled modular multiplications (see Section~\ref{sec:multiplicationgate}). 
    Our choice to simulate the multiplications using direct permutations with no extra qubits, while allowing the large-scale MPI scheme sketched in Figure~\ref{fig:mpischeme}, prevents the direct simulation of quantum errors during the multiplications (which is why, in Figure~\ref{fig:errors}, essentially only initialization errors before and measurement errors after the multiplications are shown). 
    The alternative would be to implement a general multiplication circuit using standard quantum gates and additional workspace qubits.
    To pursue this research direction to allow the study of errors during the multiplications, an informative exposition to start from is the construction by Gidney and Eker\aa~\cite{Gidney2021HowToFactor2048RSAShor}, which combines and optimizes many techniques discovered over the past decades to implement the modular multiplications.
    
    \item Although Shor's order-finding algorithm is the most prominent quantum algorithm for factoring, a practical solution of the factoring problem on gate-based quantum computers might rather use the Eker\aa-H\aa{}stad factoring scheme~\cite{Ekera2017FactoringWithDiscreteLogarithm} based on the discrete logarithm quantum algorithm (see point 3 in Section~\ref{sec:relatedwork}).
    Instead of $t\approx 2L$ stages in the iterative quantum circuit (cf.~Figure~\ref{fig:shor}) using the semiclassical Fourier transform, this algorithm requires at most $t\approx 3L/2$ stages, with a systematic option to reduce $t$ further at the cost of reducing the success probability below 99\%~\cite{Ekera2020OnPostProcessingInShor}. 
    In the context of quantum circuit simulation, the Eker\aa-H\aa{}stad scheme would save valuable execution time (cf.~Section~\ref{sec:memorycomputingtime}), allowing to gather more statistics for larger factoring scenarios.
    
    \item The \texttt{shorgpu} implementation used for this work maintains two full statevector buffers \texttt{psi} and \texttt{psibuf}, which reduces simulation time by enabling contiguous memory transfer through the MPI network (see~\cite{shorgpu}). 
    However, the total amount of memory fixes the maximum number of qubits that can be simulated, which puts a limit on the size of simulatable factoring problems.
    An alternative choice would be to use only a single statevector buffer, thereby having to replace the contiguous memory transfer with interleaved communication and computation.
    This choice (potentially combined with reducing computing time by switching to the Eker\aa-H\aa{}stad scheme, see previous item) would allow the simulation of yet another qubit, to push the boundary of simulatable factoring problems and the threshold of the proposed challenge one step further.
\end{enumerate}

\section{Conclusion}
\label{sec:4}

In this paper, we have introduced a method to simulate the iterative Shor algorithm on supercomputers with thousands of GPUs.
The simulation software~\cite{shorgpu} allowed us to push the size of factoring problems far beyond what has been achieved previously. 
We have used the simulation software
to perform an in-depth analysis of the iterative Shor algorithm.

Using Shor's original post-processing~\cite{shor1994factoring,ekert1996quantumalgorithms,shor1997algorithm,NielsenChuang}, we have shown that a significant amount of ``lucky'' factorizations raises the expected success probability from $3$--$4\,\%$ to above $50\,\%$. We have given number-theoretic arguments for the existence of the lucky cases, and we conjecture that they continue to contribute with approximately $25\,\%$ beyond the size of integer factoring problems investigated in this paper.

Using Eker\aa's post-processing~\cite{Ekera2021OnCompletelyFactoringAnyIntegerShor,Ekera2022OnTheSuccessProbabilityOfQuantumOrderFindingShor}, the success probability for a factoring scenario can be brought close to unity using only a single
bitstring obtained by executing the iterative Shor algorithm. However, Eker\aa's post-processing method assumes that the quantum part has been executed without errors, an assumption which is unlikely to hold for quantum processors in the near future. Therefore, we have studied how additional classical and quantum errors, as present in today's quantum information processing hardware~\cite{GoogleQuAI2021}, influence the performance of the post-processing procedure. Remarkably, we find that Eker\aa's post-processing procedure exhibits a particular form of universality and resilience. Here, ``universality'' means that the decrease of success probability is roughly independent of the particular type of error and ``resilience'' means that the success probability is systematically larger than the success probability expected from independent bit flip errors.

Although these results might inspire confidence in the quantum factoring procedure, the first successful factorization of a cryptographically relevant number---say RSA-2048 from the famous RSA factoring challenge---is still out of reach~\cite{Gidney2021HowToFactor2048RSAShor,Gouzien2021ShorFactoring2048RSAIn177DaysWith13436Qubits}.
Therefore, a more modest challenge towards true quantum supremacy might be 
to demonstrate that a real quantum computing device can factorize an interesting semiprime which is larger than $N_{\mathrm{max}}=549755813701$.
In fact, since gate-based quantum computers might already require full error correction for this purpose, it is conceivable  that this challenge is first met by a quantum annealer~\cite{Peng2008QuantumAdiabaticAlgorithmForFactorization,Andriyash2016BoostingIntegerFactorizationDWave,Dridi2017PrimeFactorizationDWave,Jiang2018QuantumAnnealingForPrimeFactorization,Peng2019FactoringLargeIntegersDWave,Mengoni2020BreakingRSAWithDWave2000Q,Wang2020PrimeFactorizationParemeterOptimizationIsingAnnealer}.

\section*{Author Contributions}

Conceptualization, D.W., M.W., F.J., H.D.R. and K.M.; software, D.W. and H.D.R.; validation, D.W., M.W. and H.D.R.; investigation, D.W., M.W., F.J. and H.D.R.; writing---original draft preparation, D.W.; writing---review and editing, D.W., M.W., F.J., H.D.R. and K.M.; visualization, D.W.; project administration, H.D.R. and K.M.; funding acquisition, K.M. All authors have read and agreed to the published version of the manuscript.

\section*{Funding}

The authors gratefully acknowledge the Gauss Centre for Supercomputing e.V. (www.gauss-centre.eu) for funding this project by providing computing time on the GCS Supercomputer JUWELS~\cite{JuwelsClusterBooster} at J\"ulich Supercomputing Centre (JSC).
D.W. and M.W. acknowledge support from the project J\"ulich UNified Infrastructure for Quantum computing (JUNIQ) that has received funding from the German Federal Ministry of Education and Research (BMBF) and the Ministry of Culture and Science of the State of North Rhine-Westphalia. 

\section*{Data Availability Statement}

The source code of the \texttt{shorgpu} simulator used to generate the data for this study is available at~\cite{shorgpu}.
The data that supports the findings in this study including the generated bitstrings for the individual factoring scenarios as well as further statistics are available from the corresponding author upon reasonable request.

\section*{Acknowledgments}

D.W. thanks Martin Eker{\aa{}} and Viv Kendon for helpful and stimulating discussions.
D.W. and H.D.R. thank Andreas Herten, Markus Hrywniak, and Jiri Kraus for help in optimizing the GPU-based simulation, in particular the MPI communication scheme.

\section*{Conflicts of Interest}

The authors declare no conflict of interest. The funders had no role in the design of the study; in the collection, analyses, or interpretation of data; in the writing of the manuscript; or in the decision to publish the results.

\section*{Abbreviations}
The following abbreviations are used in this manuscript:\\

\noindent 
\begin{tabular}{@{}ll}
CPU & Central Processing Unit\\
CUDA & Compute Unified Device Architecture\\
GPU & Graphics Processing Unit\\
JSC & J\"ulich Supercomputing Centre\\
JUNIQ & J\"ulich UNified Infrastructure for Quantum computing\\
JUQCS & J\"ulich Universal Quantum Computer Simulator\\
JUWELS & J\"ulich Wizard for European Leadership Science\\
MPI & Message Passing Interface\\
NISQ & Noisy Intermediate-Scale Quantum\\
QFT & Quantum Fourier Transform\\
RSA & Rivest Shamir Adleman\\
gcd & Greatest Common Divisor\\
lcm & Least Common Multiple\\
\end{tabular}

\clearpage
\onecolumngrid
\appendix

\section{The Probability Distribution Generated by Shor's Algorithm}
\label{app:distribution}

\begin{figure*}
  \centering
  \includegraphics[width=\columnwidth]{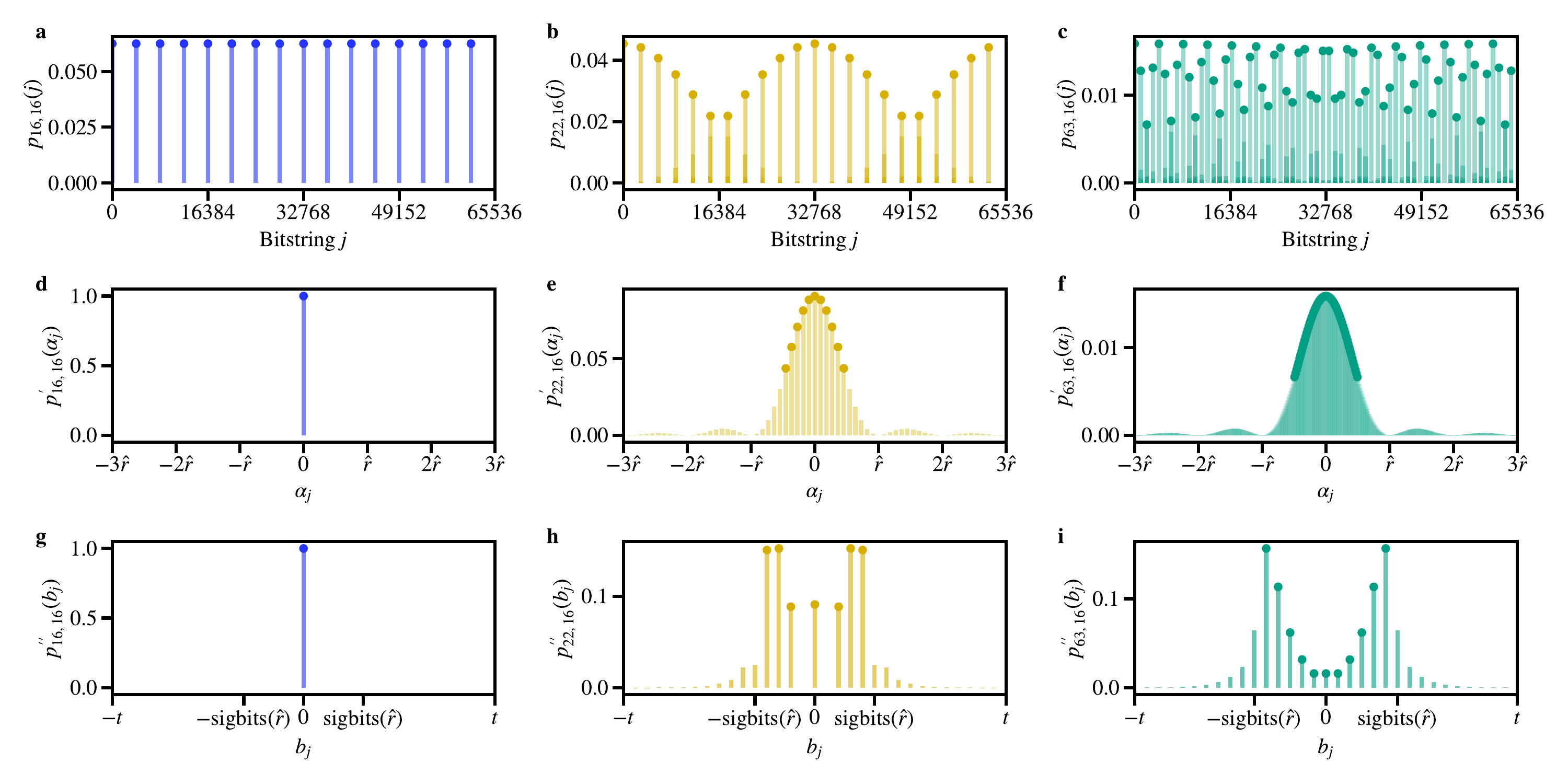}
  \caption{\textbf{Representative bitstring distributions produced by Shor's algorithm.} Shown is the probability distribution $p_{\hat r,t}(j)$ given by Eq.~(\ref{eq:distribution2}) for $t=16$ (such that the integer representation of the bitstrings $j$ ranges from $0$ to $2^t-1=65535$) and multiplicative orders \textbf{a} $\hat r=16$, \textbf{b} $\hat r=22$, and \textbf{c} $\hat r=63$. Each distribution has exactly $\hat r$ peaks (solid circles) given by Eq.~(\ref{eq:peakenumeration2}). The peaks are approximately $2^t/\hat r$ bitstrings apart. Note that in \textbf{a}, the peaks are equidistant, equally large with probability $1/\hat r$, and all other probabilities are exactly zero. These properties are lost if $\hat r$ does not divide $2^t$ evenly, as can be seen in the presence of small but non-zero bars next to the peaks in \textbf{b} and \textbf{c}.
  \textbf{d}--\textbf{f} The corresponding distributions $p_{\hat r,t}'(\alpha_j)$ expressed in terms of $\alpha_j=\{\hat rj\}_{2^t}\in\{-2^t/2,\ldots,2^t/2-1\}$ (see Eq.~(\ref{eq:probabilityalphaj})).
  \textbf{g}--\textbf{i} The corresponding distributions $p_{\hat r,t}''(b_j)$ expressed in terms of $b_j=\mathrm{sigbits}(\alpha_j)$ (see Eq.~(\ref{eq:probabilitybj})).
  }
  \label{fig:suppdistributions}
\end{figure*}

The construction of Shor's algorithm starts by assuming that there are two quantum registers of size $t$ and $L$, respectively, in the initial state $\ket{\psi_0}=\ket0\ket 0$. The first step is to bring the first register in a uniform superposition using Hadamard gates such that the state becomes
\begin{align}
    \ket{\psi_1}=\frac{1}{2^{t/2}} \sum_{k=0}^{2^t-1}\ket k\ket 0.
\end{align}
Then, application of the oracle corresponding to the function $f(k) = a^k\,\mathrm{mod}\,N$ brings the state to
\begin{align}
    \ket{\psi_2}=\frac{1}{2^{t/2}} \sum_{k=0}^{2^t-1}\ket k\ket{f(k)}.
\end{align}
The next step is to apply the quantum Fourier transform to the first register, which yields
\begin{align}
    \ket{\psi_3}=\frac{1}{2^{t}} \sum_{k=0}^{2^t-1}\sum_{j=0}^{2^t-1} e^{-2\pi i kj / 2^t} \ket j\ket{f(k)}.
\end{align}
Since $f(k)$ is a periodic function with period $\hat r$ (i.e., the multiplicative order of $a$ modulo $N$), the second register can only take $\hat r$ different values. Combining all amplitudes with equal second register gives
\begin{align}
    \ket{\psi_3}=\frac{1}{2^{t}} \sum_{k=0}^{\hat r-1}\sum_{j=0}^{2^t-1} e^{-2\pi i kj / 2^t} 
    \left(1 + e^{-2\pi i \hat rj / 2^t} + e^{-2\pi i 2 \hat rj / 2^t} + \cdots + e^{-2\pi i (s-1)\hat rj / 2^t} + e^{-2\pi i s \hat rj / 2^t}\delta_{[k+s\hat r<2^t]}\right) 
    \ket j\ket{f(k)},
\end{align}
where $s=\lfloor 2^t/\hat r\rfloor$, and $\delta_{[k+s\hat r<2^t]}$ indicates that the last term only contributes if $k+s\hat r<2^t$. Identifying a geometric sequence for the first $s$ terms, we have
\begin{align}
    \ket{\psi_3}=\frac{1}{2^{t}} \sum_{k=0}^{\hat r-1}\sum_{j=0}^{2^t-1} e^{-2\pi i kj / 2^t} 
    \left(\frac{1-e^{-2\pi is\hat rj/2^t}}{1-e^{-2\pi i\hat rj/2^t}} + e^{-2\pi i s \hat rj / 2^t}\delta_{[k+s\hat r<2^t]}\right) 
    \ket j\ket{f(k)}.
\end{align}
Finally, to obtain the probability $p_{\hat r,t}(j)$ to measure the bitstring $j$ in the first register, we trace out the second register,
\begin{align}
    p_{\hat r,t}(j) &= \frac{1}{2^{2t}} \sum_{k=0}^{\hat r-1} 
    \left\vert \frac{1-e^{-2\pi is\hat rj/2^t}}{1-e^{-2\pi i\hat rj/2^t}} + e^{-2\pi i s \hat rj / 2^t}\delta_{[k+s\hat r<2^t]}\right\vert^2 \nonumber \\
    \label{eq:distribution2}
    &= \frac{\hat r}{2^{2t}} \left( 
        \frac
        {\sin(\pi s \hat rj/2^t)}
        {\sin(\pi \hat rj/2^t)}
        \right)^2 
    + \frac{2^t - s\hat r}{2^{2t}}
        \frac
        {\sin( \pi [2s+1]\hat rj/2^t)}
        {\sin(\pi \hat rj/2^t)}.
\end{align}
This result is the same as  in~\cite{Michielsen2005EventBasedSimulationOfUniversalQC}, correcting some misprints in~\cite{einarsson2003ShorProbability} and~\cite{DeRaedt2007MassivelyParallel}.  
Note that the singularities at $\sin(\pi \hat rj/2^t)=0$ are removable singularities.

The resulting bitstring distribution is shown for a few representative cases in Fig.~\ref{fig:suppdistributions}a--c. It is strongly peaked at $\hat r$ bitstrings given by 
\begin{align}
    \label{eq:peakenumeration2}
    j\in\{0,\mathrm{round}(2^t/\hat r),\mathrm{round}(2\times2^t/\hat r),\ldots,\mathrm{round}((\hat r-1)\times2^t/\hat r)\}\;.
\end{align}
Note that when $\hat r\ge2^t$, the distribution becomes a uniform distribution that is ``peaked'' everywhere.
Furthermore, when $\hat r$ divides $2^t$ (such that $s=\lfloor2^t/\hat r\rfloor=2^t/\hat r$), only the first term in Eq.~(\ref{eq:distribution}) contributes, with the same value of $1/\hat r$ at all $\hat r$ peaks; all other bitstrings then have probability zero (see Fig.~\ref{fig:suppdistributions}a).

\subsection{Alternative Representations of the Probability Distribution}

A useful, alternative representation of the probability distribution $p_{\hat r,t}(j)$ in Eq.~(\ref{eq:distribution2}) can be obtained by identifying all bitstrings $j$ that yield equivalent arguments of the sine functions. Due to the periodicity of the sine function, these arguments can be represented by
\begin{align}
    \label{eq:alphaj}
    \alpha_j = \{\hat rj\}_{2^t} = (\hat rj + 2^t/2)\,\mathrm{mod}\,2^t - 2^t/2,
\end{align}
where the notation $\{x\}_y$ denotes $x\,\mathrm{mod}\,y$ constrained to $\{-y/2,\ldots,y/2-1\}$. 

All bitstrings $j$ that yield the same $\alpha_j$ can be enumerated by solving the equation $\alpha_j\equiv\hat r j\ (\mathrm{mod}\,2^t)$ for $j\in\{0,\ldots,2^t-1\}$. 
To do that, let $2^d$ denote the largest power of two dividing $\hat r$. Then $\hat r/2^d$ is coprime to $2^t$, so it has an inverse modulo $2^t$ which we denote by $(\hat r/2^d)^{-1}$. Thus, we find $\alpha_j (\hat r/2^d)^{-1}\equiv 2^dj\ (\mathrm{mod}\,2^t)$. This means that there is an integer $l\in\mathbb Z$ such that $2^dj=\alpha_j (\hat r/2^d)^{-1}+2^tl$.
From Eq.~(\ref{eq:alphaj}), we furthermore see that $2^d\mid\alpha_j$. Dividing by $2^d$ (note that we require $\hat r<2^t$; the other case has been discussed above) and using that $j\in\{0,\ldots,2^t-1\}$, we thus obtain
\begin{align}
    \label{eq:alphajenumerate}
    j = \left(\frac{\alpha_j}{2^d}\left(\frac{\hat r}{2^d}\right)^{-1} + 2^{t-d}l\right)\,\mathrm{mod}\,2^t.
\end{align}
Here, $l=0,\ldots,2^d-1$ enumerates all $2^d$ different bitstrings $j$.

As each $\alpha_j$ has multiplicity $2^d$ according to Eq.~(\ref{eq:alphajenumerate}), and each admissible $\alpha_j$ must be a multiple of $2^d$ according to Eq.~(\ref{eq:alphaj}), we can write the probability distribution for $\alpha_j\in\{-2^t/2,\ldots,2^t/2-1\}\cap2^d\mathbb Z$ as
\begin{align}
    \label{eq:probabilityalphaj}
    p_{\hat r,t}'(\alpha_j)
    &= 2^d\left(\frac{\hat r}{2^{2t}} \left( 
        \frac
        {\sin(\pi s \alpha_j/2^t)}
        {\sin(\pi \alpha_j/2^t)}
        \right)^2 
    + \frac{2^t - s\hat r}{2^{2t}}
        \frac
        {\sin( \pi [2s+1]\alpha_j/2^t)}
        {\sin(\pi \alpha_j/2^t)}\right).
\end{align}
This distribution is shown in Fig.~\ref{fig:suppdistributions}d--f. The first term has the typical structure of a Fraunhofer diffraction pattern. Note in particular that all peaks given by Eq.~(\ref{eq:peakenumeration2}) correspond to the values of $\alpha_j$ with $-\hat r/2\le\alpha_j\le\hat r/2$ (see~also~\cite{shor1994factoring,ekert1996quantumalgorithms,shor1997algorithm}).

The advantage of using this representation is that it is the basis of a viable method to sample from the distribution, even for cryptographically large bitstrings ~\cite{Ekera2020OnPostProcessingInShor,Ekera2021QuantumAlgorithmsWithTradeoffsSamplingShor,Ekera2021OnCompletelyFactoringAnyIntegerShor,Ekera2022OnTheSuccessProbabilityOfQuantumOrderFindingShor,Ekera2020Qunundrum}). The key is that the distribution as a function of $\alpha_j$ is quite regular and smooth, so it can be numerically integrated to obtain a cumulative distribution function. 

More precisely, one groups $\alpha_j$ into logarithmically spaced regions identified by
\begin{align}
    b_j = \mathrm{sigbits}(\alpha_j) = \begin{cases}
        \mathrm{sign}(\alpha_j)(\lfloor\log_2\vert\alpha_j\vert\rfloor+1) & (\alpha_j\neq0)\\
        0 & (\alpha_j=0)
    \end{cases},
\end{align}
which denotes the signed number of bits needed to represent the integer $\alpha_j$. This means that $2^{\vert b_j\vert-1}\le\vert\alpha_j\vert<2^{\vert b_j\vert}$ (note that for the numerical integration, one can use subregions of the form $2^{\vert b_j\vert-1+\xi/2^\nu}\le\vert\alpha_j\vert<2^{\vert b_j\vert-1+(\xi+1)/2^\nu}$~\cite{Ekera2021QuantumAlgorithmsWithTradeoffsSamplingShor}, along with Simpson's rule and Richardson extrapolation~\cite{numericalrecipes}).

The corresponding distribution,
\begin{align}
    \label{eq:probabilitybj}
    p_{\hat r,t}''(b_j)=\sum_{\mathrm{sigbits}(\alpha_j)=b_j} p_{\hat r,t}'(\alpha_j),
\end{align}
is shown in Fig.~\ref{fig:suppdistributions}g-i. The characteristic property for large $t$ is that most of the probability mass is located around $b_j\approx\pm\mathrm{sigbits}(\hat r)$. In other words, most of the sampled bitstrings $j$ have approximately as many bits as the order $\hat r$. This is independent of the particular value of $\hat r$ (unless $\hat r$ contains an artificially large power of $2$). This trend is already observable for $t=16$ in Fig.~\ref{fig:suppdistributions}i. 

In the terminology of information theory, this means that a sampled bitstring $j$ provides approximately $t-\vert\mathrm{sigbits}(\hat r)\vert$ bits of information on the order $\hat r$.
This interpretation provides another intuition for the success of Shor's algorithm. For a typical factoring problem for an $L$-bit semiprime $N=p\times q$, bitstrings with $t\approx2L$ classical bits in the recommended setting (see main text) are sampled. The multiplicative order $\hat r$ needs always \emph{less than} $L$ bits (the argument for this is that the largest possible order $\hat r$ is the least common multiple $\mathrm{lcm}(p-1,q-1)$, which is at least divisible by two, so it requires less bits than $N=p\times q$). Note that in~\cite{Ekera2021QuantumAlgorithmsWithTradeoffsSamplingShor}, the ``worst'' case that $\vert\mathrm{sigbits}(\hat r)\vert\approx L$ is considered, and even then two runs of the order-finding algorithm are sufficient.

The distribution $p_{\hat r,t}(j)$ over bitstrings $j$ with $t=16$ bits is shown in Fig.~\ref{fig:suppdistributions}. We used \texttt{shorgpu} to generate samples from the distributions $p_{\hat r,t}(j)$ with up to $t=78$ bits, without knowing the solution to the specified factoring problem. If, however, the solution to the factoring problem is known, one can use the trick explained above to generate samples of $p_{\hat r,t}(j)$ with up to $t=16384$ bits and beyond (see~\cite{Ekera2021QuantumAlgorithmsWithTradeoffsSamplingShor}).

\subsection{Probability Theory for Shor's Factoring Procedure}
\label{app:probabilitytheory}

In this section, we relate the results extracted from the large data sets to relations and theorems about Shor's algorithm found in the literature. We first reformulate the theoretical success probability for Shor's original factoring procedure in terms of probabilities for the different conditions. Then we relate each contribution to known theorems from the literature. This framework can be seen as the basis to interpret the results of Eker\aa's post-processing stated in Section~\ref{sec:postprocessingekera}.

Given an integer $N$ to factor, Shor's algorithm states that one should first pick a random $a\in\mathbb Z_N^*$ and then run the quantum algorithm. Formally, the success probability for one run of the quantum algorithm (i.e., one sampled bitstring $j$) therefore reads
\begin{align}
    p(\text{success}\mid N) 
    &= \sum_{a\in\mathbb Z_N^*} p(\text{success}\mid a,N)\:p(a\mid N).
\end{align}
We pick $a$ uniformly, so $p(a\mid N)=1/|\mathbb Z_N^*|=1/\phi(N)$, where $\phi(N)$ is Euler's totient function. Furthermore, the conditions for ``success'' stated in the literature~\cite{shor1994factoring,ekert1996quantumalgorithms,shor1997algorithm,NielsenChuang} are that the sampled bitstring $j$ yields the order $\hat r=\mathrm{ord}_N(a)$, $\hat r$ is even, and $a^{\hat r/2}\not\equiv \pm 1 \ (\mathrm{mod}\,N)$. Thus,
\begin{align}
    p(\text{success}\mid N) 
    &= \frac{1}{\phi(N)} \sum_{a\in\mathbb Z_N^*} p(\text{$j$ yields $\hat r$}\wedge\text{$\hat r$ even}\wedge a^{\hat r/2}\not\equiv_N \pm1\mid a,N).
\end{align}
We know that the bitstring $j$ yields the order $\hat r$ if $j$ is sampled at one of the $\hat k=0,\ldots,\hat r-1$ peaks of $p_{\hat r,t}(j)$ given by Eqs.~(\ref{eq:distribution2}) and (\ref{eq:peakenumeration2}), and the peak enumerator $\hat k$ is coprime to $\hat r$ (so that the continued fraction method yields the convergent $k/r=\hat k/\hat r$ with $r=\hat r$). Hence,
\begin{align}
    \label{eq:propositionsdefinition}
    p(\text{success}\mid N) 
    &= \frac{1}{\phi(N)} \sum_{a\in\mathbb Z_N^*} 
    p(\underbrace{\text{$j$ sampled at a peak}}_{\textbf{A}}
    \wedge\underbrace{\text{$\hat k$ coprime to $\hat r$}}_{\textbf{B}}
    \wedge\underbrace{\text{$\hat r$ even}\wedge a^{\hat r/2}\not\equiv_N \pm1}_{\textbf{C}}\mid a,N),
\end{align}
where we defined the propositions \textbf{A}, \textbf{B}, and \textbf{C}, the probabilities of each of which have known estimates (see below). Using the product rule~\cite{jaynes2003probability}, we have
\begin{align}
    p(\text{success}\mid N) 
    = \frac{1}{\phi(N)} \sum_{a\in\mathbb Z_N^*} 
    &p(\text{$j$ sampled at a peak}\mid a,N)
    \nonumber\\
    \vphantom{\sum_{a\in\mathbb Z_N^*}}\times\:&p(\text{$\hat k$ coprime to $\hat r$}\mid \textbf{A},a,N)
    \nonumber\\
    \label{eq:successprobability}
    \times\:&p(\text{$\hat r$ even}\wedge a^{\hat r/2}\not\equiv_N \pm1\mid \textbf{B},\textbf{A},a,N).
\end{align}
Substituting the expressions  Eqs.~(\ref{eq:propositiona}), (\ref{eq:propositionb}), and (\ref{eq:theoremProbWeak}) derived below, we arrive at the theoretical bound for the success probability,
\begin{align}
    \label{eq:finalbound}
    p(\text{success}\mid N) \gtrsim \frac{4}{\pi^2} \times\frac{e^{-\gamma}}{\log\log N}\times \frac 1 2. 
\end{align}

\begin{figure*}
  \centering
  \includegraphics[width=\columnwidth]{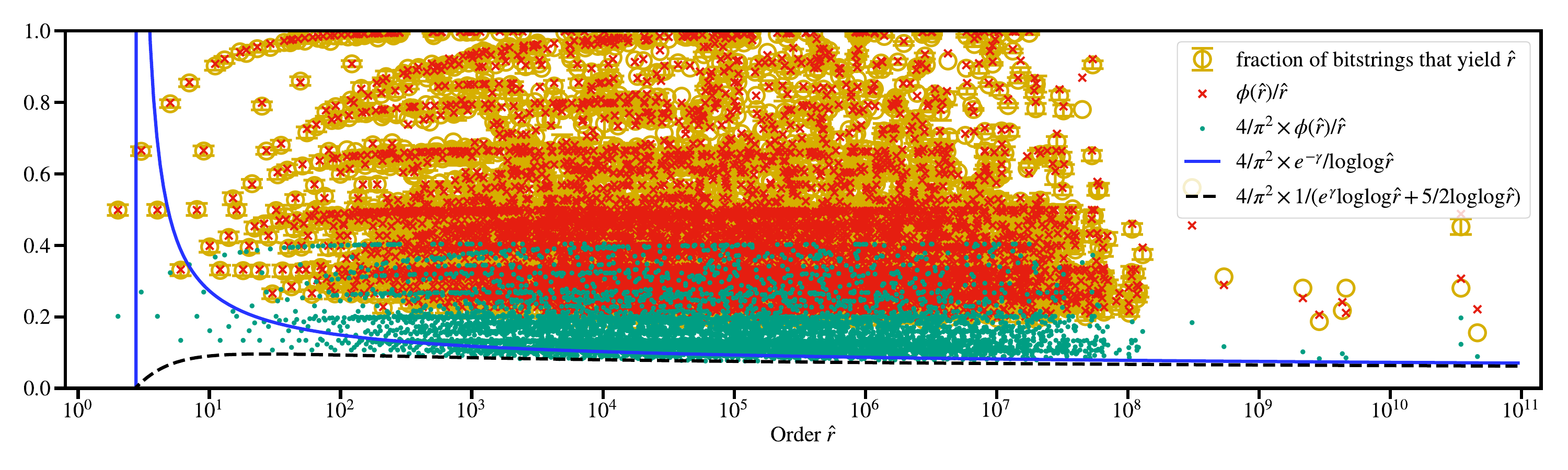}
  \caption{\textbf{Comparison of the bounds for propositions \textbf{A} and \textbf{B} proven in the literature and the corresponding frequencies extracted from the simulations.} The 52077 uniform factoring problems plus the 13 individual large cases from Fig.~\ref{fig:results}a are grouped as a function of increasing $\hat r=\mathrm{ord}_N(a)$. Yellow circles represent the average fraction of sampled bitstrings (normalized by a total of 1024 for the uniform cases and 32 for the large cases) that yield the correct order $\hat r$ (meaning that they satisfy propositions \textbf{A} and \textbf{B} in Eq.~(\ref{eq:propositionsdefinition})); error bars show the corresponding standard deviation for problems with the same $\hat r$. Red crosses indicate the corresponding values of $\phi(\hat r)/\hat r$. Green points, the solid blue line, and the dashed black line indicate the bounds in Eqs.~(\ref{eq:propositionbBound1}), (\ref{eq:boundshor}), and (\ref{eq:boundrosser}), respectively, combined with the lower bound of $4/\pi^2$ for proposition \textbf{A} in Eq.~(\ref{eq:propositiona}).
  }
  \label{fig:ordertheory}
\end{figure*}

Figure~\ref{fig:ordertheory} shows the combined bounds from propositions \textbf{A} and \textbf{B} in comparison with the corresponding data extracted from the simulations. We see that when the bound of $4/\pi^2$ for proposition \textbf{A} in Eq.~(\ref{eq:propositiona}) is included, the estimate becomes very weak. If it is not included (red crosses), the values lie only slightly above the data points (at least for all uniform factoring problems with enough samples). In other words, the probability of sampling $j$ at one of the peaks is much larger than $4/\pi^2$. 

The bound Eq.~(\ref{eq:finalbound}) takes values between $3$--$4\,\%$ for semiprimes $N$ between $2^{20}$ and $2^{40}$.
Since the actual performance of Shor's algorithm shown in Fig.~\ref{fig:results}b is clearly much better, it would be interesting to obtain better estimates and, in particular, to find statements about the averages instead of lower bounds.
\\~\\
\textbf{Proposition A:} $j\:\mathrm{sampled}\:\mathrm{at}\:\mathrm{a}\:\mathrm{peak}$
\\~\\
A known lower bound for the probability $p_{\hat r,t}(j)$ at one of the $\hat r$ peaks is $4/\pi^2$~\cite{shor1994factoring,ekert1996quantumalgorithms,shor1997algorithm}. This result can be obtained from the distribution $p_{\hat r,t}(j)$ given by 
Eq.~(\ref{eq:distribution2}): 
At a peak, we have by Eq.~(\ref{eq:peakenumeration2}) that the bitstring $j$ satisfies $j=\mathrm{round}(2^t\hat k/\hat r) = 2^t\hat k/\hat r + \delta$ with $\vert\delta\vert\le1/2$. Using $|\sin x|\ge|x|/(\pi/2)$ when $|x|\le\pi/2$, $|\sin x|\le|x|$ for all $x$, and the periodicity of $|\sin|$, we have for the numerator and the denominator of the first term,
\begin{align}
    \left|\sin(\pi j \hat r\lfloor 2^t/\hat r \rfloor/2^t)\right| &\ge 2 |\delta| \frac{\lfloor2^t/\hat r\rfloor}{2^t/\hat r} \approx 2 |\delta|,\\
    \left|\sin(\pi j \hat r/2^t)\right| &\le \pi |\delta|\frac{\hat r}{2^t}.
\end{align}
We note that the second term of $p_{\hat r,t}(j)$ in Eq.~(\ref{eq:distribution2}) is usually neglected in the literature or simply assumed to be positive. Indeed, the signs of both numerator and denominator are often dominated by the sign of $\delta$. However, it may become negative for certain values such as $\hat r=15$, $t=5$, and $j=\mathrm{round}(2^t\times4/15)$. In any case, its contribution is negligible with respect to the first term. Hence, we have $p_{\hat r,t}(j) \gtrsim 4/\pi^2\hat r$. Since there are exactly $\hat r$ peaks, we obtain
\begin{align}
    \label{eq:propositiona}
    p(\text{$j$ sampled at a peak}\mid a,N) \gtrsim \frac{4}{\pi^2} \approx 40.5\,\%.
\end{align}
We remark that considering bitstrings $j$ that are only a few steps away from a peak may also work (see~\cite{Gerjuoy2005ShorProbabilityImprovement} and also Section~\ref{sec:postprocessingekera}).
\\~\\
\textbf{Proposition B:} $\hat k\:\mathrm{coprime}\:\mathrm{to}\:\hat r$
\\~\\
The probability that an integer $\hat k=0,\ldots,\hat r-1$ is coprime to $\hat r$ is given by 
\begin{align}
  \label{eq:propositionbBound1}
  p(\text{$\hat k$ coprime to $\hat r$}\mid A,a,N) = \frac{\phi(\hat r)}{\hat r},
\end{align}
since there are exactly $\phi(\hat r)$ elements in $\mathbb Z_{\hat r}$ that are coprime to $\hat r$. There are several estimates for this quantity in the literature. Shor~\cite{shor1994factoring,shor1997algorithm} uses an estimate of the form
\begin{align}
    \label{eq:boundshor}
     \frac{\phi(\hat r)}{\hat r} \gtrsim \frac{e^{-\gamma}}{\log\log \hat r},
\end{align}
where $\gamma$ is Euler's constant such that $e^{-\gamma}\approx0.561$. This estimate is based on the fact that $\underline\lim(\phi(\hat r)\log\log \hat r/\hat r) = e^{-\gamma}$~\cite[Theorem 328]{HardyWright}.
However, this is only an infimum limit, and one can in fact show that there are infinitely many $\hat r$ violating this bound~\cite{Nicolas1983InfinitelyManyPhiThatViolateRosser}. 
Ekert and Jozsa~\cite{ekert1996quantumalgorithms} mostly argue with $\phi(\hat r)/\hat r>1/\log \hat r$ (using the prime number theorem), but this bound is only valid for $\hat r\gtrsim10^6$ and then becomes a rather weak bound. 
A better, strict lower bound to $\phi(\hat r)/\hat r$ has been proven by Rosser and Schoenfeld in~\cite{Rosser1962ApproximateFormulasPhi}, 
\begin{align}
    \label{eq:boundrosser}
     \frac{\phi(\hat r)}{\hat r} > \frac{1}{e^{\gamma}\log\log \hat r + \frac{5}{2\log\log \hat r}},
\end{align}
which is valid for all $\hat r\ge2$ except $223092870$ (in which case $5/2$ must be replaced with 2.50637).

Due to the presence of $\log\log \hat r$, both bounds in Eqs.~(\ref{eq:boundshor}) and (\ref{eq:boundrosser}) show an extremely weak dependence on $\hat r$ (e.g., for $\hat r\in\{10^2,10^{11},2^{4096}\}$, $\log\log \hat r$ varies only between 1 and 8). Therefore, either bound is suitable for the present estimate. For the same reason, we may safely approximate $\log\log \hat r\approx\log\log \phi(N)\approx\log\log N$ such that the bound becomes independent of $\hat r$. Thus, we obtain
\begin{align}
    \label{eq:propositionb}
    p(\text{$\hat k$ coprime to $\hat r$}\mid A,a,N) \gtrsim \frac{e^{-\gamma}}{\log\log N}.
\end{align}
\\~\\
\textbf{Proposition C:} $\hat r\:\mathrm{even}\wedge a^{\hat r/2}\not\equiv \pm1\ (\mathrm{mod}\,N)$
\\~\\
Combining the results for propositions \textbf{A} and \textbf{B} (which are now independent of the particular $a\in\mathbb Z_N^*$), the remaining part of Eq.~(\ref{eq:successprobability}) is
\begin{align}
    \label{eq:propositioncremaining}
    \frac{1}{\phi(N)} \sum_{a\in\mathbb Z_N^*}
    p(\text{$\hat r$ even}\wedge a^{\hat r/2}\not\equiv_N \pm1\mid B,A,a,N).
\end{align}
We note that an \emph{erroneous bound} of $1-1/2^{n_F}$ was given for this probability in both Shor's original paper~\cite{shor1994factoring} and in the book by Nielsen and Chuang~\cite{NielsenChuang}. The correct versions were given in Shor's later paper~\cite{shor1997algorithm} and in an errata list by Nielsen~\cite{NielsenChuangErrata}.
An extensive proof can be found in the review by Ekert and Jozsa~\cite{ekert1996quantumalgorithms}, which state the result as follows.
\\~\\
\textbf{Theorem:} Let $N$ be odd with prime factorization $N=p_1^{e_1}p_2^{e_2}\cdots p_{n_F}^{e_{n_F}}$. Suppose $a$ is chosen at random, satisfying $\mathrm{gcd}(a,N)=1$. Let $r$ be the order of $a\, \mathrm{mod}\,N$. Then \begin{align}
    \label{eq:theoremProbWeak}
    \mathrm{prob}(r\text{ is even and }a^{r/2}\not\equiv \pm 1 \ (\mathrm{mod}\,N)) \ge 1-\frac{1}{2^{{n_F}-1}},
\end{align}
where ``$\mathrm{prob}$'' means the frequency when enumerating all $a\in\mathbb Z_N^*$, which directly corresponds to the sum present in Eq.~(\ref{eq:propositioncremaining}). We remark that the condition $a^{r/2}\not\equiv+1$ is actually superfluous since this case does not occur if $r$ is the order (otherwise $r/2$ would already be the order).

The idea of the proof is to study the converse, namely that $r$ is odd or $a^{r/2}\equiv-1$. This only happens if all multiplicative orders $r_j$ of $a_j=a\,\mathrm{mod}\,p_j^{e_j}$ contain exactly the same power of 2 as $r$. In other words, $r/2^d$ and $r_j/2^{d_j}$ are odd integers with $d_j=d$. Summing over all possible $d$ (which may be different for different $a$) yields
\begin{align}
    \mathrm{prob}(r\text{ is odd or }a^{r/2}\equiv - 1 \ (\mathrm{mod}\,N)) = \sum_d \mathrm{prob}(d_1=d)\cdots\mathrm{prob}(d_{n_F}=d).
\end{align}
When enumerating all $a_j$, the case $d_j=d$ occurs with frequency $\le1/2$. Approximating the last ${n_F}-1$ factors $\le1/2$ and using the first factor $\mathrm{prob}(d_1=d)$ to remove the sum yields the bound $1/2^{{n_F}-1}$.

As the blue triangles in Fig.~\ref{fig:results} show, this statement is in agreement with the data, since the average of $75\,\%$ is above the bound of $50\,\%$ for ${n_F}=2$ (some error bars might extend to below $50\,\%$ which is due to the fact that we do not simulate the full set of all $a\in\mathbb Z_N^*$). Furthermore, 
the theoretical bound in Eq.~(\ref{eq:theoremProbWeak}) is also tight: For $N=21$, we have exactly $50\,\%$ of all $a\in\mathbb Z_{21}^*$ that have either an odd order $r$ or $a^{r/2}\equiv-1$. 
We do not know whether one can prove the observed average frequency of $75\,\%$ in Fig.~\ref{fig:results}b, using that $N$ is generated by uniformly drawing the prime factors $p$ and $q$ from the integers.

\section{Generation of the Factoring Problems}
\label{app:problems}

We have generated 61362 factoring problems $(N,a)$. 52077 out of these are referred to as ``uniform'' factoring problems because they have been generated by a procedure, to be described next, to ensure a uniform distribution of prime factors that is not biased towards small primes. For a given number of bits $L$, we sample the first prime factor from a uniformly distributed set of integers $p\in\{3,\ldots,\lfloor\sqrt{2^L}\rfloor\}$ until a primality test asserts that $p$ is prime. The second prime is similarly sampled from $q\in\{\lceil2^{L-1}/p\rceil,\ldots,\lfloor2^L/p\rfloor\}$ until $q>p$ and $N=p\times q$ is an $L$-bit semiprime.

We remark that the reason to consider semiprimes is that they yield the hardest factoring problems when factoring is reduced to order finding. This is because many elements in $\mathbb Z_N^*$ have large orders, but the largest order $\lambda(N)$ (i.e., the Carmichael function~\cite{Carmichael1910}) is always less than $N/2^{n_F-1}$, where $n_F\ge2$ is the number of distinct prime factors in $N$~\cite[Claim 7]{Ekera2022OnTheSuccessProbabilityOfQuantumOrderFindingShor}. Thus, if $N$ has more than $n_F=2$ factors, the orders become smaller on average and thus easier to find.

For each $L=4,\ldots,28$, this procedure is done for 50 different $N$. For each $N$, we subsequently draw 50 different $a\in\{2,\ldots,N-1\}$ coprime to $N$. This procedure exhausts all $N$ for $4\le N\le8$ and generates 2500 unique problems $(N,a)$ for each $L>8$. For each problem, \texttt{shorgpu} generated $M=1024$ bitstrings.

In addition to the uniform factoring problems, we generated 9285 individual problems relevant for Figs.~\ref{fig:results}a, \ref{fig:scalet}, and \ref{fig:errors}. In particular, these problems include the individual ``large'' cases with $30\le n \le 40$ qubits, for which we always choose the largest interesting semiprimes $N$ (see Table~\ref{tab:largestsemiprimes} in Appendix~\ref{app:tables}). The number of sampled bitstrings is $M=32$ ($M=128$) for the results presented in Fig.~\ref{fig:results}a (Fig.~\ref{fig:errors}a). In case none of these bitstrings yields a factor, we continue with a second random $a$. This is the reason that for $n=31$ and $n=37$, one pair of ``success'' and ``success$+$lucky'' markers is at $0\,\%$ and only the second pair is above $0\,\%$.

\section{Standard Procedure: Shor's Post-Processing}
\label{app:standardprocedure}

Executing Shor's algorithm for a given factoring problem $(N,a)$ yields a bitstring $j$ with $t$ bits (the recommended number of bits is $t=\lceil2\log_2N\rceil$, which comes from the requirement that $N^2\le2^t<2N^2$~\cite{shor1997algorithm}; so we always have $t\in\{2L,2L-1\}$ since $N$ is no power of two).
From the continued fraction expansion of $j/2^t$ (using the integer representation of the bitstring $j$), one takes the convergent $k/r$ with the largest denominator $r<N$~\cite{shor1994factoring,ekert1996quantumalgorithms,shor1997algorithm} (we remark that in principle, it is better to stop at the largest denominator $r< 2^{t/2}$, otherwise one can construct pathological examples for smaller $t$ for which going up to $N$ skips the order and yields an unrelated, larger integer; see also~\cite[Lemma 6]{Ekera2022OnTheSuccessProbabilityOfQuantumOrderFindingShor}). The resulting $r$ is often (cf.~Fig.~\ref{fig:histograms}b) equal to the order $\hat r$ of $a$ modulo $N$, i.e., the smallest exponent such that $a^{\hat r}\,\mathrm{mod}\,N=1$. The standard procedure dictates that if $r$ is even, $a^{r}\,\mathrm{mod}\,N=1$, and $a^{r/2}\,\mathrm{mod}\,N\neq-1$, then computing the greatest common divisors $\mathrm{gcd}(a^{r/2}\pm1,N)$ has a high probability of yielding a factor of $N$. Recall that in this work, if one of these checks on $r$ fails but $\mathrm{gcd}(a^{\lfloor r/2\rfloor}\pm1,N)$ still produces a factor, the bitstring $j$ is counted as ``lucky''.

\section{List of Semiprimes}
\label{app:tables}

In Table \ref{tab:largestsemiprimes}, we give a list of the largest interesting semiprimes with $L<50$ bits, for which a factorization using the iterative Shor algorithm shown in Fig.~\ref{fig:shor} would need up to $n=50$ qubits.

\begin{table}[p!]
  \caption{List of the largest interesting semiprimes (where ``interesting'' means that the two prime factors are distinct and have the same number of decimal digits) that can be factored using the iterative Shor algorithm for a given number of qubits $n=L+1$, where $L$ is the number of bits required to represent the semiprime. For each semiprime $N=p\times q$, $t=\lceil \log_2 N^2\rceil$ is the recommended minimum number of classical bits to read out (cf.~Fig.~\ref{fig:shor}).}
  \begin{center}
    \begin{ruledtabular}
      \begin{tabular}{rrrrr}
       qubits $n$ & semiprime $N$ & factor $p$ & factor $q$ & $t$ \\
        \colrule
5 & 15 & 3 & 5 & 8 \\
6 & 21 & 3 & 7 & 9 \\
7 & 35 & 5 & 7 & 11 \\
8 & 35 & 5 & 7 & 11 \\
9 & 253 & 11 & 23 & 16 \\
10 & 493 & 17 & 29 & 18 \\
11 & 1007 & 19 & 53 & 20 \\
12 & 2047 & 23 & 89 & 22 \\
13 & 4087 & 61 & 67 & 24 \\
14 & 8051 & 83 & 97 & 26 \\
15 & 16241 & 109 & 149 & 28 \\
16 & 32743 & 137 & 239 & 30 \\
17 & 65509 & 109 & 601 & 32 \\
18 & 131029 & 283 & 463 & 34 \\
19 & 262099 & 349 & 751 & 36 \\
20 & 524137 & 557 & 941 & 38 \\
21 & 1048351 & 1009 & 1039 & 40 \\
22 & 2097101 & 1399 & 1499 & 42 \\
23 & 4194163 & 1307 & 3209 & 44 \\
24 & 8388563 & 2357 & 3559 & 46 \\
25 & 16777207 & 4093 & 4099 & 48 \\
26 & 33554089 & 3797 & 8837 & 50 \\
27 & 67108147 & 8011 & 8377 & 52 \\
28 & 134217449 & 11119 & 12071 & 54 \\
29 & 268435247 & 12589 & 21323 & 56 \\
30 & 536870861 & 22717 & 23633 & 58 \\
31 & 1073741687 & 27779 & 38653 & 60 \\
32 & 2147483551 & 32063 & 66977 & 62 \\
33 & 4294967213 & 57139 & 75167 & 64 \\
34 & 8589933181 & 89597 & 95873 & 66 \\
35 & 17179869131 & 125627 & 136753 & 68 \\
36 & 34359737977 & 117517 & 292381 & 70 \\
37 & 68719476733 & 242819 & 283007 & 72 \\
38 & 137438953319 & 189853 & 723923 & 74 \\
39 & 274877906893 & 364303 & 754531 & 76 \\
40 & 549755813701 & 712321 & 771781 & 78 \\
41 & 1099511623591 & 1002817 & 1096423 & 80 \\
42 & 2199023255179 & 1286533 & 1709263 & 82 \\
43 & 4398046510399 & 2014013 & 2183723 & 84 \\
44 & 8796093021439 & 2217443 & 3966773 & 86 \\
45 & 17592186044353 & 2005519 & 8771887 & 88 \\
46 & 35184372088787 & 3769453 & 9334079 & 90 \\
47 & 70368744177439 & 8388593 & 8388623 & 92 \\
48 & 140737488355141 & 11150957 & 12621113 & 94 \\
49 & 281474976708763 & 15847327 & 17761669 & 96 \\
50 & 562949953421083 & 16619039 & 33873797 & 98 \\ 
\end{tabular}
      \end{ruledtabular}
    \label{tab:largestsemiprimes}
  \end{center}
\end{table}

\clearpage
\bibliographystyle{apsrev4-2custom}
\bibliography{bibliography}

\end{document}